\documentclass[a4paper,10pt]{article}
\usepackage{amsmath,amssymb,epsfig,graphicx} 
\usepackage{setspace}

\begin{document}


\title{Consistent simulation of non-resonant diphoton production 
at hadron collisions with a custom-made parton shower}

\author{Shigeru Odaka\footnote{E-mail: \texttt{shigeru.odaka@kek.jp}.}, 
and Yoshimasa Kurihara\\
High Energy Accelerator Research Organization (KEK)\\
1-1 Oho, Tsukuba, Ibaraki 305-0801, Japan}

\date{}

\maketitle


\begin{abstract}
We have developed a Monte Carlo event generator for non-resonant diphoton 
($\gamma\gamma$) production at hadron collisions in the framework of GR@PPA, 
which consistently includes additional one-jet production.
The jet-matching method developed for initial-state jet production 
has been extended to the final state in order to regularize the final-state 
QED divergence in the $qg \rightarrow \gamma\gamma + q$ process.
A QCD/QED-mixed parton shower (PS) has been developed to complete the matching.
The PS has the capability of enforcing hard-photon radiation, 
and small-$Q^{2}$ photon radiations that are not covered by the PS are 
supplemented by using a fragmentation function. 
The generated events can be passed to general-purpose event generators 
in order to perform the simulations down to the hadron level.
Thus, we can simulate the isolation requirements that must be 
applied in experiments at the hadron level.
The simulation results are in reasonable agreement with the predictions 
from RESBOS and DIPHOX.
The simulated hadron-level events can be further fed to detector simulations 
in order to investigate the detailed performance of experiments.
\end{abstract}

\section{Introduction}

Diphoton ($\gamma\gamma$) production is one of the most promising channels 
for the discovery of the Higgs boson having a relatively small invariant mass 
($\lesssim 140$ GeV/$c^{2}$) at the CERN LHC. 
However, the measurements suffer from the large irreducible diphoton background 
produced by non-resonant electromagnetic (QED) 
interactions~\cite{ATLAS:2012sk,CMS:2012tw}.
It is necessary to understand the properties of this background, 
not only for the search for the Higgs boson but also for detailed studies 
after the discovery.
A precise understanding cannot be achieved without detailed knowledge 
of the photon-identification capabilities of experiments.
In hadron-collision experiments, 
photon identification is influenced by soft hadrons that are 
additionally produced by various soft phenomena, such as initial- and 
final-state radiations, hadronization, decays, underlying events 
and event pile-ups,  
because a certain isolation condition has to be required in order 
to reduce the large contamination of $\pi^{0}$ from hadron jets. 
The effects of such soft phenomena cannot be reliably evaluated 
by analytical calculations or parton-level simulations.
Hence, it is strongly desired to provide theoretical predictions 
in the form of Monte Carlo (MC) event generators 
with which we can carry out simulations down to the hadron level.

\begin{figure}[t]
\centerline{\psfig{file=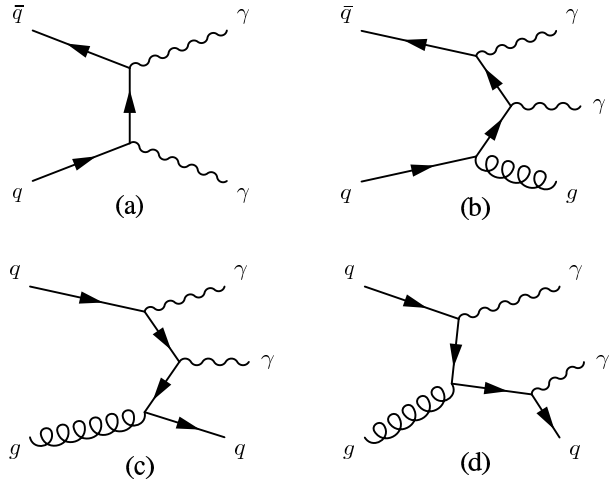,width=100mm}}
\caption{Typical Feynman diagrams for non-resonant diphoton production 
at hadron collisions: 
(a) the lowest order, (b) a gluon radiation process, 
(c) a quark radiation process, and (d) another quark radiation process.
Processes (b) and (c) have initial-state QCD divergences, 
while (d) has a final-state QED divergence.
\protect\label{fig:diagrams}}
\end{figure}

The lowest-order process for non-resonant diphoton production is very simple, 
as shown in Fig.~\ref{fig:diagrams}(a). 
Despite this, the next-to-leading order (NLO) correction to this process is 
known to be very large~\cite{Binoth:1999qq,Balazs:2007hr}.
The large correction is predominantly due to the contribution from real 
radiation processes, illustrated in Figs.~\ref{fig:diagrams}(b) to (d).
While the gluon-radiation contribution from Fig.~\ref{fig:diagrams}(b) 
should not be very large, the contribution from quark-radiation 
processes shown in Figs.~\ref{fig:diagrams}(c) and (d) may become large 
due to the large gluon density inside protons.
It is necessary to include these processes in order to provide 
a realistic simulation.
However, available MC event generators consistently supporting 
these processes are limited
because of a difficulty in treating the divergences associated 
with photon radiation.
Although a recent release of SHERPA~\cite{Hoeche:2009xc} seems to support 
such photon production, 
no other MC event generators are available as far as we know\footnote{
A forthcoming release of HERWIG++~\cite{Bahr:2008pv} may provide a consistent 
simulation as an NLO event generator~\cite{D'Errico:2011sd}.}. 

In this report, we describe an MC event generator for non-resonant QED 
diphoton production in hadron collisions that we have developed, 
and discuss its predictions.
The program has been developed in the framework of the GR@PPA event 
generator~\cite{Odaka:2011hc,Odaka:2012da} 
and supports the generation of radiative processes 
in Figs.~\ref{fig:diagrams}(b) to (d), 
which include one additional jet (light quark or gluon) in the final state,
together with the lowest-order process in Fig.~\ref{fig:diagrams}(a).
The matrix elements (ME) of these processes are generated 
using the GRACE system~\cite{Ishikawa:1993qr}.
Though the event generator includes radiative processes, 
it is not fully including NLO corrections 
because non-divergent terms in soft/virtual corrections are yet to be included.
In any case, the radiative processes have various divergences 
which we need to regularize.
The initial-state strong-interaction (QCD) divergences can be regularized 
using a method that we have developed for weak-boson production processes, 
where divergent components are numerically subtracted from the matrix elements 
of radiative processes 
(the LLL subtraction)~\cite{Kurihara:2002ne,Odaka:2007gu,Odaka:2009qf}.
The subtracted components are restored by combining non-radiative processes 
to which a parton shower (PS) is applied.
We can avoid the double-count problem by the subtraction and 
regularize the divergences as a result of the multiple radiation in PS.

The quark-radiation processes illustrated in Figs.~\ref{fig:diagrams}(c) and (d) 
has not only an initial-state QCD divergence but also a QED divergence 
in the final state.
We have extended the method for initial-state QCD divergences 
to this final-state QED divergence.
The extension of the subtraction method is straightforward, 
while the preparation of an appropriate PS is not trivial 
since it has to support QED together with QCD.
This PS has to be applied to the $qg \rightarrow \gamma q$ process 
to radiate photons from the final-state quark based on 
the collinear approximation (the fragmentation process).
We have successfully developed such a PS on the basis of the QCD PS 
included in the GR@PPA 2.8 distribution\footnote{The base release is 
the GR@PPA 2.8.3 update~\cite{Odaka:2012da}.}.

Since parton showers are based on perturbative calculations, 
it is necessary to apply a certain cutoff to avoid collinear 
(small-$Q^{2}$) divergences.
Further soft phenomena in QCD including non-perturbative effects 
are simulated by adopting appropriate models in general-purpose event 
generators, {\it e.g.}, PYTHIA~\cite{Sjostrand:2006za}.
However, such simulations are not available for QED photon radiation.
Since we consider radiations from final-state partons, 
energetic photons visible in detectors can emerge even 
from very small-$Q^{2}$ branches.
In order to simulate such small-$Q^{2}$ photon radiations, 
we have adopted a method employing a fragmentation function (FF).
The PS supplemented with an FF-based simulation that we have developed 
has the capability of enforcing hard-photon radiation.
This function dramatically improves the generation efficiency 
for the fragmentation process.

The generated events can be passed to PYTHIA~\cite{Sjostrand:2006za} 
for simulating small-$Q^{2}$ QCD phenomena down to the hadron level.
Hence, we can evaluate the effect of isolation requirements 
at the hadron level using the simulated events.
Those events processed by PYTHIA can be fed to 
detector simulations of experiments.
Further detailed studies will become possible with such simulations.

We sometimes find reports in which the detection efficiency and acceptance 
for diphoton measurements are evaluated by using event generators 
for the lowest-order process and the fragmentation process.
Such evaluations are not self-consistent.
Parton showers used in the simulation of the fragmentation process have 
an energy scale to be determined arbitrarily. 
The energy scale defines the maximum hardness of the radiation.
The simulation results depend on this energy scale 
since those radiations exceeding the scale are ignored.
Non-collinear contributions are also ignored.
These ignored contributions may be small in many processes compared to 
the contributions taken into consideration.
However, as we will show in this report, they may become comparable to 
the lowest-order contribution in diphoton production. 
It is necessary to include a simulation based on exact matrix elements 
for radiative processes in order to make a reliable evaluation.

The fragmentation process that we take into consideration in this report 
is the so-called "one-fragmentation" process.
The "two-fragmentation" process in which two photons are radiated 
from final-state partons, for instance, 
from quarks in $gg \rightarrow q\bar{q}$, is not supported.
Besides, the gluon-fusion process, $gg \rightarrow \gamma\gamma$, and 
its higher orders are not included in the present study.

We require a typical kinematical condition for the Higgs-boson search 
at the LHC through the present study 
because we are interested in its background; that is
\begin{eqnarray}
  & p_{T}(\gamma_{1}) \geq 40 \text{ GeV/}c, \ 
  p_{T}(\gamma_{2}) \geq 25 \text{ GeV/}c, \nonumber \\
  & |\eta(\gamma)| \leq 2.5, \ \Delta R(\gamma\gamma) \geq 0.4 \nonumber \\
  & 80 \leq m_{\gamma\gamma} \leq 140 \text{ GeV/}c^{2}.
\label{eq:selection}
\end{eqnarray}
We apply an asymmetric requirement to the transverse momenta ($p_{T}$) of 
photons with respect to the incident beam direction.
The requirement on the pseudorapidity ($\eta$) is common to the two photons.
In addition, though this is not effective for real diphoton events 
now we consider, 
we require a sufficient $\Delta R$ separation between the two photons, 
where $\Delta R$ is defined from the differences in the pseudorapidity ($\eta$) 
and azimuthal angle ($\phi$) as 
$\Delta R^{2} = \Delta\eta^{2} + \Delta\phi^{2}$.
Finally, the invariant mass of the two photons ($m_{\gamma\gamma}$) is 
restricted to the range that we are interested in.
The simulations are carried out for the design condition of the LHC, 
proton-proton collisions at a center-of-mass (cm) energy of 14 TeV, 
with CTEQ6L1~\cite{Pumplin:2002vw} for the parton distribution function (PDF).

This report is organized as follows: 
the extension of the limited leading-log (LLL) subtraction method to the 
final-state QED divergence is described in Section 2, 
and the QCD/QED-mixed PS that we have developed is described in Section 3.
The matching in the final-state QED radiation is discussed in Section 4.
Results from a full simulation down to the hadron level are presented 
and the effect of a typical isolation requirement is discussed in Section 5.
Physical results obtained from the full simulation are presented 
and discussed in Section 6, 
and the discussions are concluded in Section 7.

\section{Final-state QED LLL subtraction}

We approximate the final-state QED divergence in the matrix element for 
the $qg \rightarrow \gamma\gamma + q$ process as
\begin{equation}
  \left|\mathcal{M}_{qg \rightarrow \gamma\gamma q}^{(\text{LLL,fin})}
  (\hat{s}, {\hat \Phi}_{\gamma\gamma q})\right|^{2}
  = \left|\mathcal{M}_{qg \rightarrow \gamma q}(\hat{s}_{0}, 
  {\hat \Phi}_{\gamma q})\right|^{2}
  f^{(\text{LL,fin})}_{q \rightarrow q \gamma}(Q^{2},z) 
  \ g(\hat{s},\hat{s}_{0})  
  \ \theta(\mu_{\text{LLL}}^{2} - Q^{2}) .
\label{eq:lll}
\end{equation}
The leading-log (LL) radiation function can be given as 
\begin{equation}\label{eq:rad}
  f^{(\text{LL,fin})}_{q \rightarrow q \gamma}(Q^{2},z)={\alpha \over 2\pi}  
  {16 \pi^{2} \over Q^{2}} P_{q \rightarrow q \gamma}(z)
\end{equation}
with the splitting function that is defined as 
\begin{equation}\label{eq:split}
  P_{q \rightarrow q \gamma}(z) = e_{q}^{2} {1+z^{2} \over 1-z},
\end{equation}
where $e_{q}$ is equal to 1/3 for down-type quarks and 2/3 for up-type quarks.
The electromagnetic coupling, $\alpha$, is assumed to be constant 
and equal to 1/132.51, 
the value used for matrix-element calculations~\cite{Tsuno:2006cu}.
We evaluate Eq.~(\ref{eq:lll}) for two possible combinations of 
$q$-$\gamma$ pairs in the final state, 
and numerically subtract them from the matrix element of 
$qg \rightarrow \gamma\gamma q$, 
together with the LLL term for the initial-state QCD 
divergence~\cite{Odaka:2011hc}.
The parameters $Q^{2}$ and $z$ are defined by using the sum of the energy 
($E_{q\gamma}$) and momentum ($p_{q\gamma}$) of the considered 
$q$-$\gamma$ pair.
They are defined in the cm frame of the $qg \rightarrow \gamma\gamma q$ 
event as 
\begin{equation}\label{eq:qsq}
  Q^{2} = E_{q\gamma}^{2} - p_{q\gamma}^{2},
\end{equation}
and
\begin{equation}\label{eq:z}
  z = { p_{L}( E_{q\gamma} + p_{q\gamma}) + Q^{2}/2 \over p_{q\gamma}
  ( E_{q\gamma} + p_{q\gamma}) + Q^{2} },
\end{equation}
where $p_{L}$ is the momentum component of the $q$ 
in parallel to the summed momentum.
Equation~(\ref{eq:z}) is defined so that $z$ should represent the momentum 
fraction at the infinite-momentum limit, 
$Q^{2}/p_{q\gamma}^{2} \rightarrow 0$~\cite{Odaka:2011hc}.

Equation\ (\ref{eq:rad}) is slightly different from the radiation function for 
the initial state \cite{Kurihara:2002ne}; $z$ is absent in the denominator.
This factor is included in the initial-state radiation function in order to take 
into account the change of the flux factor in the cross-section calculation, 
which is equal to the squared cm energy of the considered process.
The relation between the squared cm energy of the non-radiative subprocess 
($\hat{s}_{0}$) and that of the radiative process ($\hat{s}$) can be 
expressed as $\hat{s}_{0} = z\hat{s}$ for initial-state radiations, 
whereas final-state radiations do not alter the cm energy at the limit 
$Q^{2}/p^{2} \rightarrow 0$ where theories are usually discussed.
Therefore, if we naively follow this argument, 
we should set $\hat{s}_{0} = \hat{s}$ and $g(\hat{s},\hat{s}_{0}) = 1$ 
in Eq.~(\ref{eq:lll}). 

In our PS model, the kinematics of PS branches are so defined that 
the momentum is strictly conserved whereas the energy conservation 
is ignored~\cite{Odaka:2012da}.
The energy conservation is restored by increasing the initial-state parton 
momenta after completing the PS simulation.
The mapping from a $qg \rightarrow \gamma\gamma q$ event to a non-radiative 
$qg \rightarrow \gamma q$ event, 
which is necessary for evaluating the LLL term in Eq.~(\ref{eq:lll}), 
has to strictly reverse this procedure in order to achieve a good matching.
We define the quark momentum of the non-radiative process as the sum of 
the momenta of the quark and one of the two photons in the radiative process.
Hence, for non-zero $Q^{2}$, $\hat{s}_{0}$ is necessarily smaller 
than $\hat{s}$.
Accordingly, the flux factor also changes in the final-state radiation.
In order to take this change into account, we define the correction factor 
in Eq.~(\ref{eq:lll}) as
\begin{equation}\label{eq:corr_lll}
 g(\hat{s},\hat{s}_{0}) = \hat{s}/\hat{s}_{0} .
\end{equation}
We note that this kind of correction, which is effective only at large $Q^{2}$, 
strongly depends on the applied PS model.
The correction in the above must be appropriate only for the PS model 
that we have adopted.

\begin{figure}[t]
\centerline{\psfig{file=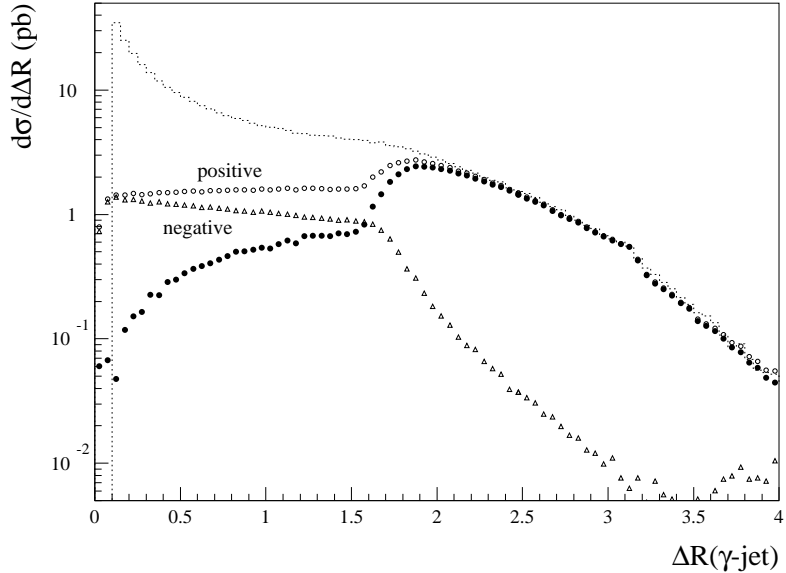,width=120mm}}
\caption{$\Delta R(\gamma\text{-jet})$ distribution of simulated 
$qg \rightarrow \gamma \gamma + q$ events 
satisfying the kinematical condition in Eq.\ (\ref{eq:selection}).
The dotted histogram shows the distribution before applying the final-state 
QED LLL subtraction, 
where $\Delta R(\gamma\text{-jet}) > 0.1$ is required to cut off the divergence.
The open circles and triangles show the distribution of positive- and 
negative-weight events, respectively, after the subtraction. 
The final result after applying the final-state subtraction is shown 
with filled circles.
The initial-state QCD LLL subtraction has already been applied 
while parton showers are yet to be applied in this simulation.
\protect\label{fig:drgamq}}
\end{figure}

An event generator implementing the above subtraction has been developed 
in the framework of GR@PPA, 
and the generation was tested for the 14-TeV LHC condition.
The energy scales were defined as~\cite{Odaka:2011hc}
\begin{equation}\label{eq:scale}
  \mu^{2} = | \vec{p}_{T}(\gamma_{1}) - \vec{p}_{T}(\gamma_{2}) |^{2}/4,
\end{equation}
where $\vec{p}_{T}(\gamma_{i})$ denotes the transverse momentum vector 
of the two photons.
This definition is equivalent to the ordinary definition, 
$\mu = p_{T}(\gamma)$, for $q\bar{q} \rightarrow \gamma\gamma$ events.
We used the identical definition for the renormalization and factorization scales.
The energy scale for the initial-state PS must be equal to the factorization 
scale in our method.
Though it is not necessary, 
we adopted the same definition for the final-state PS.
The energy scale to limit the LLL subtraction, 
$\mu_{\rm LLL}$ in Eq.~(\ref{eq:lll}), 
must be equal to the energy scale of the final-state PS to be applied to 
the mapped non-radiative event~\cite{Odaka:2012da}.
We defined it to be equal to the $p_{T}$ of the mapped 
$qg \rightarrow \gamma q$ event in the current study.

A result is shown in Fig.~\ref{fig:drgamq}.
We have plotted the distribution of $\Delta R$ between the photon and quark 
in the final state of the $qg \rightarrow \gamma \gamma + q$ events, 
where $q$ represents any quark or anti-quark up to the $b$ quark.
We obtain two values since there are two photons in the final state.
We take the smaller one as $\Delta R(\gamma\text{-jet})$.
The constraint on the photons in Eq.~(\ref{eq:selection}) was required 
at the event generation.
Parton showers are yet to be applied but the initial-state QCD LLL subtraction 
has been applied in this simulation.
Therefore, the result is finite even though we have required practically 
nothing for the final-state quark. 
The quark is allowed to be very soft and out of the detection.
This is why we denote the process as $qg \rightarrow \gamma \gamma + q$ 
instead of $qg \rightarrow \gamma \gamma q$.
The distribution without the application of the final-state LLL subtraction 
is shown by the dotted histogram for comparison,
in which we have applied a cutoff of $\Delta R(\gamma\text{-jet}) > 0.1$ 
in order to avoid the divergence at $\Delta R(\gamma\text{-jet}) = 0$.

Since the LLL subtraction is unphysical, we obtain negative-weight events 
as well as ordinary positive-weight events when we apply the subtraction.
The event weights are always equal to $+1$ or $-1$ 
because BASES/SPRING~\cite{Kawabata:1985yt,Kawabata:1995th} automatically 
unweights the events.
Therefore, we can obtain the desired distribution by subtracting the number 
of negative-weight events from that of positive-weight events 
in each histogram bin.
The open circles and triangles in the figure show the distribution of 
positive- and negative-weight events, respectively, 
and the final distribution is shown by filled circles. 
We can see that the distributions after the subtraction converge to 
finite values as $\Delta R(\gamma\text{-jet}) \rightarrow 0$;
not only does the final result converge to zero, but the distributions of 
positive- and negative-weight events individually converge to a finite value.
These facts imply that the subtraction was performed properly, 
at least near the divergence limit.
We have applied a very small cutoff, $\Delta R(\gamma\text{-jet}) > 0.01$, 
in the event generation for numerical stability.
We can see that the effect of this cutoff is negligible.

The LLL subtraction is limited by the $\theta$ function in Eq.~(\ref{eq:lll}).
A steep rise in the final distribution between 
$\Delta R(\gamma\text{-jet}) = 1.5$ and 2.0 
corresponds to this limitation.
The subtraction is not effective in the large $\Delta R(\gamma\text{-jet})$ 
region where events composed of two photons widely separated 
from the final-state quark exhibit.
The distribution below the rise shows a non-logarithmic contribution 
remaining after the subtraction.
The distribution of negative-weight events steeply decreases at large 
$\Delta R(\gamma\text{-jet})$ since they emerge only in the region 
where the subtraction is active. 
The negative-weight events at very large $\Delta R(\gamma\text{-jet})$ 
were produced by the initial-state QCD LLL subtraction.

\section{Fragmentation process}

The subtracted LL components have to be restored by non-radiative processes 
to which an appropriate PS is applied.
The initial-state QCD LL components in $q\bar{q} \rightarrow \gamma\gamma + g$ 
and $qg \rightarrow \gamma\gamma + q$ can be restored with 
$q\bar{q} \rightarrow \gamma\gamma$ events to which an ordinary initial-state 
QCD PS is applied, as in the case of weak-boson production.
On the other hand, the final-state QED LL components in 
$qg \rightarrow \gamma\gamma + q$ have to be restored with 
$qg \rightarrow \gamma q$ events to which a final-state PS including 
QED photon radiations is applied.
We have developed such a PS.

\subsection{QCD/QED-mixed PS}

The implementation of parton showers is very simple.
First of all, we define the Sudakov form factor which provides the probability 
that no branch happens between certain energy scales.
It can be expressed by using the radiation probability, $\Gamma_{i}(Q^{2})$, 
as
\begin{equation}\label{eq:sudakov}
  S(Q_{1}^{2}, Q_{2}^{2}) = \exp\left[ - \int_{Q_{1}^{2}}^{Q_{2}^{2}}
  dQ^{2}\sum_{i}\Gamma_{i}(Q^{2}) \right] ,
\end{equation}
where the sum is taken over all relevant branch modes.
The radiation probability can be obtained by integrating the splitting 
probability, $\Psi(Q^{2},z)$, over the "momentum fraction" $z$ as
\begin{equation}\label{eq:radpro}
  \Gamma_{i}(Q^{2}) =  \int_{z_{{\rm min},i}}^{z_{{\rm max},i}} 
  \Psi_{i}(Q^{2},z) dz .
\end{equation}
Although the integration should be done from 0 to 1 ideally, 
we must frequently limit the range in order to regularize the divergences 
in the splitting probabilities.
In our QCD PS~\cite{Odaka:2011hc}, 
the splitting probability is approximated as 
\begin{equation}\label{eq:splitpro}
  \Psi_{i}(Q^{2},z) =  {1 \over Q^{2}}\ {\alpha_{s}(Q^{2}) \over 2\pi} P_{i}(z) 
\end{equation}
by using the 1-loop QCD coupling, $\alpha_{s}(Q^{2})$, and the leading-order 
splitting functions, $P_{i}(z)$.

In the actual implementation of PS, 
we first determine the $Q^{2}$ of a branch.
Suppose that we have a branch at $Q^{2} = Q_{\rm pre}^{2}$.
In the final-state (timelike) PS, 
the $Q^{2}$ of the next branch can be determined by solving the equation, 
$S(Q^{2}, Q_{\rm pre}^{2}) = \eta$, by introducing a random number $\eta$ 
which uniformly distributes in the range from 0 to 1.
Once the $Q^{2}$ is determined, 
we can determine the branch mode $i$ according to the radiation probability 
$\Gamma_{i}(Q^{2})$, 
and then determine the $z$ parameter in proportion to $\Psi_{i}(Q^{2},z)$.
The thus determined PS parameters, $Q^{2}$ and $z$, are converted to the momenta 
of participating partons based on a certain kinematics model. 
The model that we have adopted is discussed 
elsewhere~\cite{Odaka:2011hc,Odaka:2012da,Odaka:2007gu}.

We have extended the above discussion to QED.
The Sudakov form factor and the radiation probability 
for QED radiations of quarks can be defined in the same way as in the above 
by using the QED splitting probability defined as
\begin{equation}\label{eq:splitpro_qed}
  \Psi_{\rm QED}(Q^{2},z) =  {1 \over Q^{2}}\ {\alpha \over 2\pi} 
  P_{q \rightarrow q \gamma}(z) .
\end{equation}
The splitting function $P_{q \rightarrow q \gamma}(z)$ is defined 
in Eq.~(\ref{eq:split}).
We can consider this QED branch as one of the branch modes for defining 
the total Sudakov form factor in Eq.~(\ref{eq:sudakov}), 
and apply the same procedure for determining the PS parameters, 
$Q^{2}$ and $z$.
Although it is not straightforward to solve the equation 
$S(Q^{2}, Q_{\rm pre}^{2}) = \eta$ because the $Q^{2}$ dependence is 
different in the QCD and QED Sudakov form factors, 
we can obtain an accurate solution by performing an iteration.
We found that if an appropriate iteration is applied we can achieve 
an accuracy better than $10^{-10}$ for the $S(Q^{2}, Q_{\rm pre}^{2})$ 
value with three or four iteration steps.

The above is a straightforward extension.
The so-called "old model" of the PYTHIA PS~\cite{Sjostrand:2006za}, 
PYSHOW, employs a different method for generating QED branches.
The $Q^{2}$ of the next branch is determined independently for possible 
QCD and QED branches, 
and the branch giving a larger $Q^{2}$ is taken as the next one.
Although the implementation is very different, 
these two methods give an identical probability for QED radiation.
When we have a branch at $Q^{2} = Q_{\rm pre}^{2}$, 
the probability that a QED branch at $Q^{2} = Q_{\rm QED}^{2}$ 
is the next branch is 
$S(Q_{\rm QED}^{2}, Q_{\rm pre}^{2})\Gamma_{\rm QED}(Q_{\rm QED}^{2})$
for the both methods, 
where $S(Q_{\rm QED}^{2}, Q_{\rm pre}^{2})$ is the total Sudakov form factor 
between the two energy scales and
$\Gamma_{\rm QED}(Q_{\rm QED}^{2})$ is the QED radiation probability 
at $Q^{2} = Q_{\rm QED}^{2}$.
We have confirmed that these two methods give an identical result 
using our simulation program.
Since the result is identical, 
it is better to use the PYSHOW method because it is faster; 
the iteration in the first method takes time.

A mixed PS which equally treats QCD and QED branches has been 
constructed by adding the above procedure to our PS.
Incidentally, the implementation of PS is terminated at a certain $Q^{2}$ 
($= Q_{0}^{2}$).
We set $Q_{0} = 5 {\rm ~GeV}$ in our PS.
As has been discussed in the Introduction, 
hard photons can be radiated with even smaller $Q^{2}$.
In order to simulate such photon radiations, 
we add QED branches to partons remaining after the PS implementation 
according to the FF in Ref.~\cite{Bourhis:1997yu}. 
We use the source code included in the DIPHOX 1.2 
distribution~\cite{Binoth:1999qq}.
Although this FF is based on the next-to-leading-log (NLL) approximation, 
and thus the gluons may radiate photons, 
we add the branches to the remaining quarks only.
This FF program has two options.
The difference between the two options predominantly results in
different radiation probabilities from gluons.
Since we do not consider the radiation from gluons, 
the difference is not significant in our application.
We adopt the first option in the current study.

The FF gives us the probability of the photon radiation having 
a certain momentum fraction of $x$.
The probability is integrated over the $Q^{2}$ range from 
a given energy scale ($Q_{\rm FF}^{2}$) down to, in principle, $Q^{2} = 0$.
We set $Q_{\rm FF}^{2} = Q_{0}^{2}$. 
Though the FF gives us longitudinal momentum information, 
it does not provide any information on the transverse motion.
However, this is not a serious problem in our application 
because the relevant $Q^{2}$ values are always small.
Although it may be sufficient to add $q \rightarrow q\gamma$ branches 
without any transverse motion, we add a non-zero $p_{T}$ to the branch products 
with respect to the parent-parton direction.
The $p_{T}$ value is randomly chosen according to a Gaussian distribution. 
We set the standard deviation of the Gaussian distribution to 
1 GeV/$c$ tentatively in the present study.

\subsection{Forced photon-radiation PS}

A mixed PS supplemented with an FF radiation has been successfully 
constructed as described in the previous subsection.
However, since the QED coupling is markedly smaller than the QCD coupling, 
it is not efficient to generate hard photons that we are interested in 
by using such a PS.
It is desired to introduce a mechanism to enforce hard-photon radiation.

First of all, we consider the probability 
that the first QED branch is at $Q^{2} = Q_{\rm QED}^{2}$.
The probability of there being a QCD branch at $Q^{2} = Q_{\rm pre}^{2}$ 
is $\Gamma_{\rm QCD}(Q_{\rm pre}^{2})$.
The probability of a QED branch at $Q^{2} = Q_{\rm QED}^{2}$ 
being the next branch is 
$S(Q_{\rm QED}^{2}, Q_{\rm pre}^{2})\Gamma_{\rm QED}(Q_{\rm QED}^{2})$, 
as described in the previous subsection.
Of course, there should be no QED branch from a given PS energy scale 
($\mu^{2}$) down to $Q_{\rm pre}^{2}$.
Therefore, the probability that we want to evaluate can be expressed as
\begin{eqnarray}\label{eq:qedpro1}
  &&P_{\rm QED}^{(1)}(Q_{\rm QED}^{2},\mu^{2}) = 
  S(Q_{\rm QED}^{2},\mu^{2})\Gamma_{\rm QED}(Q_{\rm QED}^{2}) \nonumber\\
  &&+ \int_{Q_{\rm QED}^{2}}^{\mu^{2}} dQ_{\rm pre}^{2} 
  S_{\rm QED}(Q_{\rm pre}^{2},\mu^{2}) \times \Gamma_{\rm QCD}(Q_{\rm pre}^{2})
  \times S(Q_{\rm QED}^{2}, Q_{\rm pre}^{2})\Gamma_{\rm QED}(Q_{\rm QED}^{2}) .
\end{eqnarray}
We have added the first term in order to include the probability 
that the QED branch is the very first branch.
Since $S(Q_{1}^{2}, Q_{2}^{2}) = S_{\rm QCD}(Q_{1}^{2}, Q_{2}^{2}) 
S_{\rm QED}(Q_{1}^{2}, Q_{2}^{2})$, 
$S(Q_{1}^{2}, Q^{2}) S(Q^{2}, Q_{2}^{2}) = S(Q_{1}^{2}, Q_{2}^{2})$ and  
\begin{equation}
  {d \over dQ_{2}^{2}} S_{i}(Q_{1}^{2}, Q_{2}^{2}) = 
  - S_{i}(Q_{1}^{2}, Q_{2}^{2}) \Gamma_{i}(Q_{2}^{2}) ,
\end{equation}
the expression in Eq.~(\ref{eq:qedpro1}) can be converted to 
\begin{equation}\label{eq:qedpro2}
  P_{\rm QED}^{(1)}(Q_{\rm QED}^{2},\mu^{2}) = 
  S_{\rm QED}(Q_{\rm QED}^{2},\mu^{2}) \Gamma_{\rm QED}(Q_{\rm QED}^{2}) .
\end{equation}
This is the probability of finding the first QED branch at 
$Q^{2} = Q_{\rm QED}^{2}$ without taking QCD branches into consideration.
Namely, the $Q^{2}$ of the first QED branch can be determined 
independently of QCD. 
We have verified this conclusion using our simulation.

A PS which enforces hard-photon radiation has been constructed based on 
the above discussion.
We consider only one photon-radiation from each prompt quark produced 
by the hard interaction,
and do not take into account the radiation from secondary quarks 
which are produced from gluon splitting.
This assumption must be sufficient because the photon-radiation probability 
is very small.
The FF photon radiation is also considered for the primary quarks only.
In addition, we apply a constraint that the radiated photons always have 
large energy ($\geq E_{\rm min}$).
The requirement of a hard-photon radiation results in certain constraints 
on random numbers used for executing the PS.
The constraints determine the weight of the generated event.
If this event weight is fed to the BASES/SPRING system appropriately, 
BASES/SPRING can automatically perform the cross-section integration and 
event generation according to the constraints that we require.
The actual implementation of our PS is described in the following.
The application of this PS improved the event-generation efficiency 
by a factor of 88 in the study to be described later.

Before starting the event generation,  
we evaluate the total FF-radiation probability for each quark flavor.
We can evaluate it since $Q_{\rm FF}$ is always equal to the $Q_{0}$ of PS.
When a quark to which we should apply the PS is selected, 
we can evaluate the total PS photon-radiation probability from 
the Sudakov form factor $S_{\rm QED}(Q_{0}^{2},\mu^{2})$.
The selection of the quark is trivial in the present study 
because each event includes only one quark in the final state.
We estimate the total photon-radiation probability by adding 
the FF and PS probabilities,
and take this probability as the base event-weight.
Then, we determine whether the photon should be radiated from FF or PS 
according to the evaluated probabilities.
If PS is chosen, 
we determine $Q_{\rm QED}^{2}$ at which the QED branch should happen 
by solving the equation $S_{\rm QED}(Q_{\rm QED}^{2}, \mu^{2}) = \eta$.
The random number $\eta$ is constrained in the range to give a solution 
within $Q_{0}^{2} < Q_{\rm QED}^{2} < \mu^{2}$, 
because this constraint is already taken into account in the event weight.

The QCD PS is applied as usual after the above procedure.
During the evolution, 
when $Q^{2}$ for the next branch becomes smaller than 
$Q_{\rm QED}^{2}$, a QED branch is inserted instead of a QCD branch.
Of course, the insertion is done only if the PS radiation is selected.
Then, we determine the $z$ parameter of the QED branch.
We apply a hard-photon constraint here.
Though the full three-dimensional kinematics is yet to be determined, 
the energy of the branching parton is available 
at this stage of the simulation.
Thus, the requirement of $E_{\gamma} \geq E_{\rm min}$ can be converted to 
an allowed range of the $z$ parameter.
This constraint results in an additional event weight.
The weight is zero 
if the energy of the branching parton is not large enough.
After completing the determination of the $z$ parameter, 
the QCD evolution is restarted by setting the maximum $Q^{2}$ 
to $Q_{\rm QED}^{2}$.
After completing the PS simulation, a photon radiation is added 
according to the FF if the FF radiation is selected.
The hard-photon requirement gives another event weight here 
because it constrains the allowed range of the $x$ parameter in the FF.

The total event weight is evaluated as the product of all event weights, 
and passed to BASES/SPRING using the LabCut framework 
which is experimentally implemented in GR@PPA 2.8. 
Parton showers can be applied in both the integration 
and event-generation stages of BASES/SPRING in this framework.
A cut can be applied to the events after the PS simulations.
This framework is useful for generating events in which the kinematics 
of the particles of interest are strongly affected by PS, 
such as photon production with relatively small-$p_{T}$, 
because in principle we can apply a very loose condition 
at the hard-interaction generation.
Incidentally, no cut is applied in the present study; 
the LabCut framework is used only for passing the event weight 
to BASES/SPRING.
We have confirmed that the constructed PS (the forced PS) gives 
results identical to those from the mixed PS 
above the hard-photon requirement, $E_{\gamma} \geq E_{\rm min}$.

\subsection{Corrections}

The kinematics model which determines the relation between the PS parameters, 
$Q^{2}$ and $z$, and the parton momenta in the PS branch
is described elsewhere~\cite{Odaka:2011hc,Odaka:2012da}.
The model is designed so that the application of the final-state PS should 
never alter the momenta of the other particles in the final state.
The application of a final-state PS necessarily increases the total energy 
in the final state 
since it makes on-shell partons develop into massive jets.
The energy conservation is restored by increasing the initial-state parton 
momenta in our model.
In the course of the present study, 
we found that this model is too naive to properly simulate the particle 
momenta inside jets.
The implementation of PS is theoretically justified only at the collinear 
limit where the increase of the energy can be ignored.
The implementation may substantially depend on the applied model away from 
the limit.
Our PS model increases the squared cm energy of the hard process, $\hat{s}$, 
and the cross sections of the hard processes that we are now interested in 
commonly decrease in proportion to $1/\hat{s}$.
Hence, it must be natural to suppress the PS implementation 
in proportion to $1/\hat{s}$.
In addition, since we increase the initial-state parton momenta, 
we should also take the change in the parton distribution function (PDF) 
into consideration. 

In order to take these $\hat{s}$ effects into account, 
we apply a correction to the generated events as follows.
The correction factors are evaluate as event weights.
The weight relevant to the cross-section change is defined as
\begin{equation}\label{eq:corr-shat}
  w_{\hat{s}} = \hat{s}/\hat{s}' ,
\end{equation}
where $\hat{s}$ and $\hat{s}'$ represent the squared cm energy before 
and after the application of the final-state PS, respectively.
The initial-state parton momenta are increased so that this increase should 
not alter the rapidity of the hard-interaction system.
Hence, the momentum fraction of the initial-state partons is adjusted as 
$x_{i}' = \sqrt{\hat{s}'/\hat{s}}x_{i}$ 
where $i$ represents two initial-state partons (1 or 2).
The weight corresponding to the change of the parton density is defined as
\begin{equation}\label{eq:corr-pdf}
  w_{\rm PDF} = { f_{1}(x_{1}',\mu_{F}^{2})f_{2}(x_{2}',\mu_{F}^{2}) \over 
  f_{1}(x_{1},\mu_{F}^{2})f_{2}(x_{2},\mu_{F}^{2}) } ,
\end{equation}
where $f_{i}(x_{i},\mu_{F}^{2})$ is the PDF for the parton $i$.
We assume that the factorization scale, $\mu_{F}^{2}$, is unchanged.
The total weight is evaluated as $w = w_{\hat{s}}w_{\rm PDF}$.
Since this weight is always smaller than one, 
we can apply a correction with a simple hit/miss rejection.
If the event is rejected,
the final-state PS is retried until the event is accepted. 
Therefore, this correction never alters the cross section of the event 
nor the momenta of final-state particles to which the PS is not applied.
The correction solely alters the momentum distribution of particles 
inside PS and the resultant jet-mass distribution.
Of course, the correction vanishes at the collinear limit.

\subsection{Simulation result}

\begin{figure}[tp]
\centerline{\psfig{file=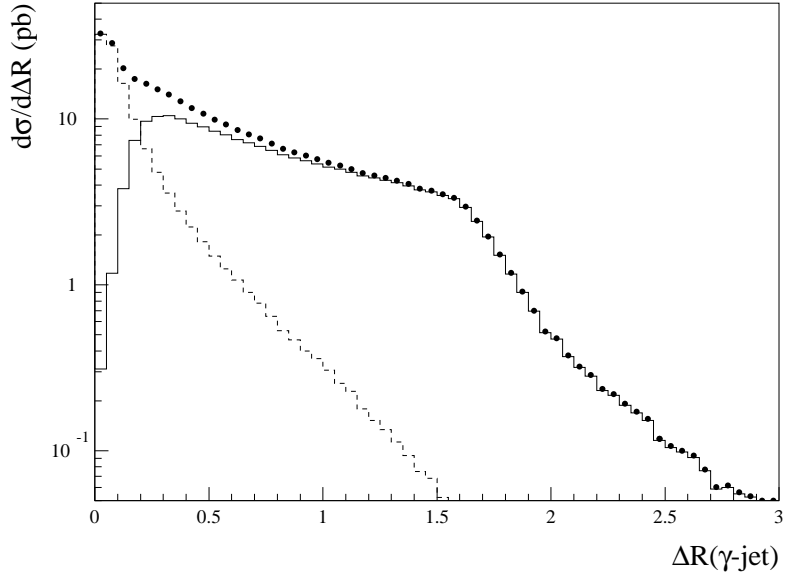,width=120mm}}
\caption{$\Delta R(\gamma\text{-jet})$ distribution of the fragmentation process. 
The forced PS with a constraint of $E_{\gamma} \geq 20 {\rm ~GeV}$ has been 
applied to the $qg \rightarrow \gamma q$ events generated in the LHC condition.
The initial-state PS is not applied.
The kinematical condition in Eq.~(\ref{eq:selection}) is required 
after the PS simulation.
Together with the total yield (plots), 
the PS (solid histogram) and FF (dashed histogram) contributions are 
separately shown.
\protect\label{fig:frag}}
\end{figure}

The $\Delta R(\gamma\text{-jet})$ distribution of the fragmentation process 
simulated with the developed PS is shown in Fig.~\ref{fig:frag}.
The forced photon-radiation PS described in the above has been applied to 
the $qg \rightarrow \gamma q$ events generated in the LHC condition. 
The PS energy scale, $\mu_{\rm FSR}$, was chosen to be equal to the $p_{T}$ 
of the generated events. 
The generation condition was sufficiently relaxed in order not to affect 
the final result.
The initial-state PS is yet to be applied.
We required a constraint of $E_{\gamma} \geq 20 {\rm ~GeV}$ in the forced PS.
The kinematical condition in Eq.~(\ref{eq:selection}) was required 
after the PS simulation.
Thus, the constraint in the forced PS does not affect the obtained result.

In order to derive the quantity $\Delta R(\gamma\text{-jet})$, 
we need to determine the momentum of the remnant jet in which the detected 
fragmentation photon is excluded.
We define it from the initial-state parton momenta and the observed two 
photon momenta, without using the momenta of partons produced by PS. 
Although the obtained jet momentum is equal to the summed momentum 
of remnant partons in our simulation, 
this definition can be universally applied to other simulations 
in which the origin of final-state partons may be ambiguous.
In any case, by using the obtained remnant-jet momentum, 
we can reconstruct a $qg \rightarrow \gamma\gamma + q$ event 
which can be compared with the parton-level $qg \rightarrow \gamma\gamma + q$ 
events generated according to the matrix elements.
Since the jet mass of the remnant jet is ignored, 
the cm energy of the reconstructed $qg \rightarrow \gamma\gamma + q$ events 
is smaller than that of the simulated events.
The energy conservation is restored by adjusting the initial-state parton 
momenta as is done in the PS application.

In Fig.~\ref{fig:frag}, 
together with the total yield (plots), 
we have illustrated the PS (solid histogram) and FF (dashed histogram) 
contributions separately.
The FF contribution is about 40\% of the total, 
and is concentrated at small $\Delta R(\gamma\text{-jet})$. 
An unnatural bumpy structure at small $\Delta R(\gamma\text{-jet})$ 
evident in the total distribution 
suggests that the transverse motion added to FF branches must be too small. 
However, we do not need to be greatly concerned about this detail because 
$\Delta R(\gamma\text{-jet})$ is a quantity which is hard to accurately 
measure in actual experiments, 
especially at small $\Delta R(\gamma\text{-jet})$ 
where the fragmentation photon is confined inside a hadron jet.
The tail in the FF contribution can be attributed to an effect of multiple 
QCD branches in PS.
The distribution at large $\Delta R(\gamma\text{-jet})$ is dominated by 
the PS contribution. 
The distribution is strongly suppressed above 
$\Delta R(\gamma\text{-jet}) = 1.5$ 
owing to the limitation by $\mu_{\rm FSR}$.

\section{Matching test}

\begin{figure}[tp]
\centerline{\psfig{file=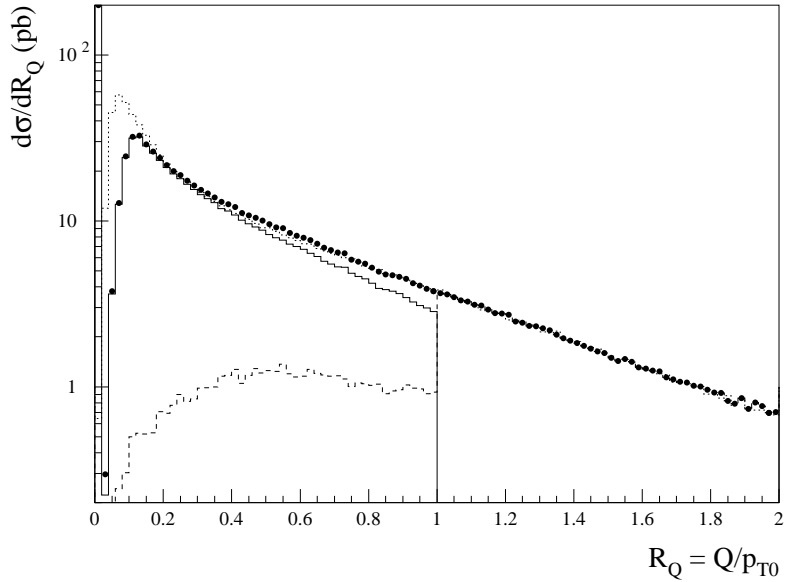,width=120mm}}
\caption{$R_{Q}$ distributions of simulated fragmentation events 
(solid histogram) and LLL-subtracted $qg \rightarrow \gamma \gamma + q$ 
events (dashed histogram).
The sum of the two distributions is illustrated as a plot.
See the text for the definition of the variable.
A naive prediction from the $qg \rightarrow \gamma \gamma + q$ ME 
with a constraint of $\Delta R(\gamma {\rm -jet}) > 0.1$ is shown 
with a dotted histogram for comparison.
\protect\label{fig:matching}}
\end{figure}

The simulation of the $qg \rightarrow \gamma \gamma + q$ process is 
completed by adding the two simulations, 
the LLL-subtracted $qg \rightarrow \gamma \gamma + q$ process 
and the fragmentation process.
The matching between the two simulations can be rigorously tested 
at the parton level by using the distribution of a parameter 
$R_{Q} = Q/p_{T0}$. 
In the fragmentation events, 
$Q$ is defined as the square root of the $Q^{2}$ parameter of the PS branch 
in which the fragmentation photon has been produced, 
and the $p_{T}$ of the prompt photon is taken as $p_{T0}$.
Thus, $Q$ is not an observable quantity.
Observable kinematical variables are affected by multiple QCD radiations 
in the fragmentation events.
Such an alternation is a higher-order effect 
which is not taken into account in the parton-level 
$qg \rightarrow \gamma \gamma + q$ events.
We can perform an unambiguous test at a common perturbation order 
by using the thus-defined quantity $Q$.

In our PS model, the parameter $Q$ is assumed to be 
the invariant mass of the branch products. 
Hence, in the LLL-subtracted $qg \rightarrow \gamma \gamma + q$ events, 
the quantity that should be compared with $Q$ is the invariant 
mass of a $\gamma$-$q$ pair.
There are two possible values since two photons exist in the final state.
We take the smaller one as $Q$, and take the $p_{T}$ of the unpaired 
photon as $p_{T0}$. 
This $p_{T}$ is the same as $p_{T}$ of the non-radiative subprocess assumed 
in the LLL subtraction.
Because we take the $p_{T}$ of the non-radiative process as the final-state 
PS energy scale, $\mu_{\rm FSR}$, in the fragmentation events, 
and as the upper limit of the subtraction, $\mu_{\rm LLL}$, 
in the LLL-subtracted events by default, 
the LL component must be sharply shared between the two simulations at 
$R_{Q} = 1$ in the default setting.

The simulation results are shown in Fig.~\ref{fig:matching}.
The distributions were obtained with the simulations 
described in previous sections.
The distribution of the fragmentation events is shown with a solid histogram, 
and that of the LLL-subtracted $qg \rightarrow \gamma \gamma + q$ events with 
a dashed histogram.
The dashed histogram at $R_{Q} < 1$ shows the remaining non-LL component 
in the $qg \rightarrow \gamma \gamma + q$ process.
We set $Q = 0$ for FF branches.
Hence, the FF-radiation events in the fragmentation process are 
concentrated in the first bin.
The distribution for the fragmentation process that is extended 
up to $R_{Q} = 1$ is determined by PS branches.

The two simulation results have sharp edges at $R_{Q} = 1$ as expected.
The sum of the two distributions is plotted with filled circles.
We can see that the two simulations are connected smoothly at the boundary, 
and that the summed distribution has a natural shape without any remarkable 
discontinuity.
The naive prediction from the $qg \rightarrow \gamma \gamma + q$ ME 
with a constraint of $\Delta R(\gamma\text{-jet}) > 0.1$
is drawn with a dotted histogram for comparison.
The summed distribution is in good agreement with this prediction down to 
$R_{Q} \sim 0.3$.
The simulation result has to be suppressed with respect to the naive 
ME-prediction at small $R_{Q}$ 
since the former is finite whereas the latter is divergent.
We find that the suppression is effective only below $R_{Q} \sim 0.3$.

The agreement between the simulation and the ME prediction 
at $R_{Q} > 1$ is trivial since no subtraction is applied in this region.
On the other hand, the agreement at $R_{Q} < 1$ is not trivial because 
the fragmentation result, which dominates the distribution in this region, 
depends on the applied corrections.
The contribution from the non-LL component is also significant.
The marked difference between the PS distribution and the ME prediction is 
completely supplemented by the non-LL contribution.
We have confirmed that we always find a significant mismatch if any of the 
corrections described in previous sections is disregarded.
Incidentally, the matching is not perfect.
A small enhancement with respect to the ME prediction 
can be seen in the $R_{Q}$ range from 0.3 to 0.7.
The enhancement is at most 10\% and far smaller than 
the higher-order QCD effects to be described later.

\begin{figure}[tp]
\centerline{\psfig{file=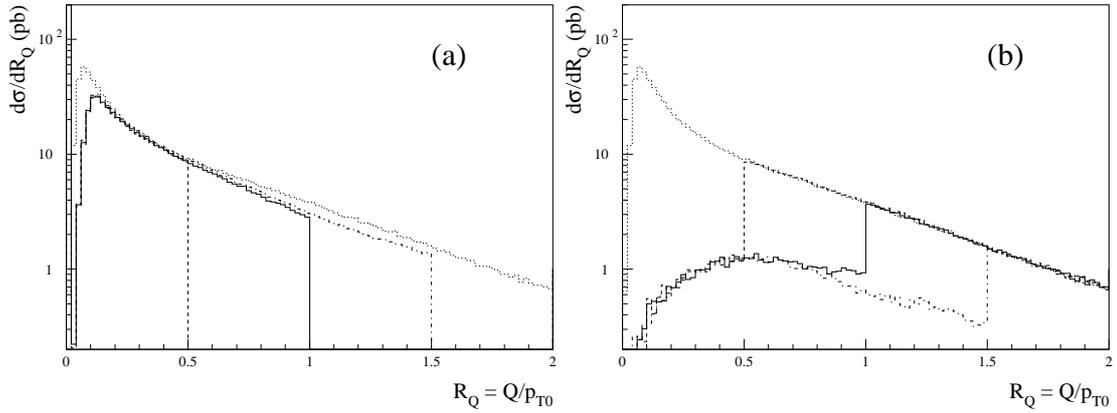,width=160mm}}
\caption{Energy-scale dependence of the $R_{Q}$ distributions 
for the fragmentation process (a) and 
the LLL-subtracted $qg \rightarrow \gamma \gamma + q$ process (b).
The results are shown for three settings: 
$\mu/p_{T0} =$ 0.5 (dashed), 1.0 (solid) and 1.5 (dot-dashed), 
where $\mu = \mu_{\rm FSR} = \mu_{\rm LLL}$.
The naive ME-prediction with $\Delta R(\gamma {\rm -jet}) > 0.1$ is shown 
with a dotted histogram for comparison.
\protect\label{fig:scale}}
\end{figure}

\begin{figure}[tp]
\centerline{\psfig{file=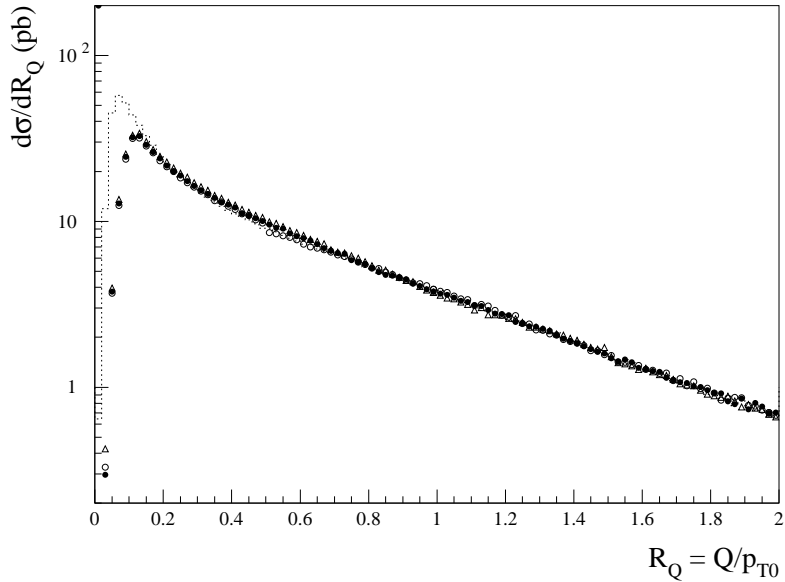,width=120mm}}
\caption{Energy-scale dependence of the $R_{Q}$ distribution.
The sum of the distributions in Fig.~\ref{fig:scale} is plotted for 
three settings for the energy scale: $\mu/p_{T0} =$ 0.5 (open circles), 
1.0 (filled circles) and 1.5 (open triangles).
The dotted histogram shows the naive ME-prediction with 
$\Delta R(\gamma {\rm -jet}) > 0.1$.
\protect\label{fig:scale_sum}}
\end{figure}

We set $\mu_{\rm FSR} = \mu_{\rm LLL} = p_{T0}$ in the above study.
This choice is arbitrary.
If the matching is perfect, the result should be unchanged 
even if we take other settings.
We have tested the stability by changing the definition as 
$\mu = \mu_{\rm FSR} = \mu_{\rm LLL} = 0.5 p_{T0}$ and $1.5 p_{T0}$.
The boundary in the $R_{Q}$ distribution moves according to this change, 
as shown in Fig.~\ref{fig:scale}.
The sum of the distributions is over-plotted in Fig.~\ref{fig:scale_sum} for 
the three settings.
Despite the fact that the small enhancement at medium $R_{Q}$ that we found 
in Fig.~\ref{fig:matching} produces a small mismatch 
in the case of $\mu = 0.5 p_{T0}$, 
the three distributions are hard to distinguish from each other.
Namely, the distribution is very stable against the change of the energy scale.
The summed cross section is also stable within $\pm$6\%.

\begin{figure}[tp]
\centerline{\psfig{file=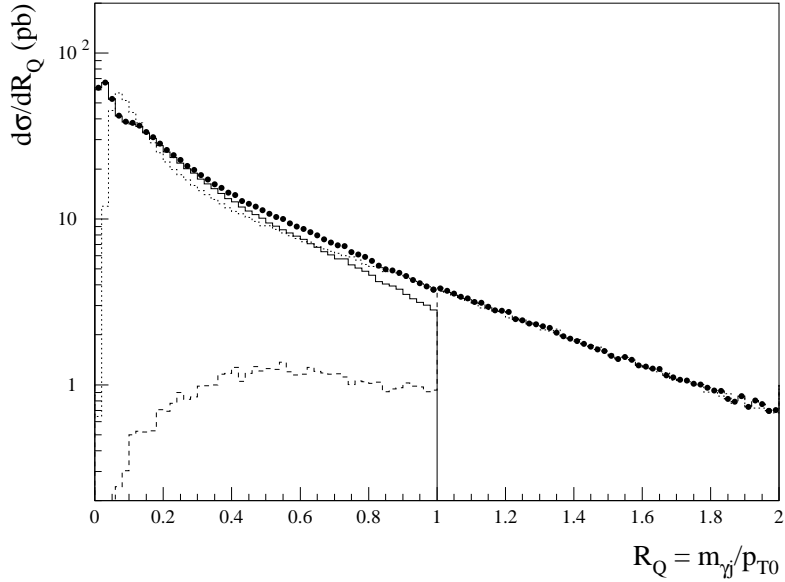,width=120mm}}
\caption{As for Fig.~\ref{fig:matching}, but $Q$ for the fragmentation 
events is defined as the invariant mass between the photon 
and quark ($m_{\gamma j}$) in the reconstructed parton-level 
$qg \rightarrow \gamma \gamma + q$ events.
\protect\label{fig:matching2}}
\end{figure}

The matching study described above has almost nothing to do with 
the kinematics model in the PS.
The model only indirectly contributes to the tested distribution through 
the $\hat{s}$ corrections.
Figure~\ref{fig:matching2} shows another $R_{Q}$ distribution, 
in which $R_{Q}$ for the fragmentation events is defined using 
the reconstructed parton-level $qg \rightarrow \gamma \gamma + q$ events 
in the same way as in the LLL-subtracted $qg \rightarrow \gamma \gamma + q$ 
events.
Hence, the distribution should reflect the effect of the kinematics model.
As we can see in Fig.~\ref{fig:matching2}, the change of the definition 
does not very significantly alter the distribution. 
The distribution for the LLL subtracted $qg \rightarrow \gamma \gamma + q$ 
is the same as that shown in Fig.~\ref{fig:matching}.
The FF radiations contribute to the distribution at small $R_{Q}$ 
in this result.

The matching is still good, 
although we can see a significant enhancement in a wide range of $R_{Q}$ 
with respect to the naive ME-prediction, 
and a sight discontinuity at $R_{Q} = 1$.
These alternations can be attributed to the effects of QCD radiations 
preceding the QED radiation.
The previously used $Q$ must be identical to the $\gamma q$ 
invariant mass in our model
if the QED branch is the first branch in PS.
Preceding QCD branches smears this identity.
The smearing results in the enhancement in the $R_{Q}$ distribution 
because the distribution is a decreasing function of $R_{Q}$.
Although they are not visible in Fig.~\ref{fig:matching2}, 
a very small but substantial amount of fragmentation events exhibit 
even at $R_{Q} > 1$ as a result of the smearing.
This leakage causes the discontinuity at $R_{Q} = 1$.
Since the smearing is a higher-order effect which is not included 
in the lowest-order $qg \rightarrow \gamma \gamma + q$ ME, 
these alternations should be considered as an advantage of our simulation. 
A similar enhancement in the weak-boson $p_{T}$ spectrum at medium $p_{T}$ 
is important to precisely reproduce the measurement data~\cite{Odaka:2009qf}.
Note that the final-state QCD radiation effects are not taken into account 
in RESBOS and DIPHOX.

\section{Full simulation and isolation cut}

It is necessary in actual experiments to require an isolation condition 
in the identification of photons 
in order to reduce the huge background from hadron jets.
It is difficult to reproduce the cuts applied by experiments with parton-level 
simulations.
This is the main reason why hadron-level event generators are desired.
Besides, if hadron-level events are generated consistently, 
the generated events can be passed to detector simulations for more detailed 
studies.

In this section, we simulate a typical isolation condition by using events 
which have been simulated down to the hadron level.
The event generation was carried out in the 14-TeV LHC condition separately 
for the fragmentation process and the other processes including the 
$q\bar{q} \rightarrow \gamma\gamma$, 
$q\bar{q} \rightarrow \gamma\gamma + g$, 
and $qg \rightarrow \gamma\gamma + q$ processes, 
where $q$ denotes any quark and anti-quark up to the $b$ quark.
The latter was generated simultaneously 
with the initial-state QCD and final-state QED LLL subtractions activated 
for radiative processes which include a jet ($q$ or $g$) in the final state. 
The initial- and final-state parton showers (PS) were fully applied 
down to $Q_{0} = 5.0 {\rm ~GeV}$ in both event generations, 
where a backward-evolution PS was used for the initial-state PS.
The forced photon-radiation PS supplemented with the FF radiations was 
applied to the fragmentation process with a constraint of 
$E_{\gamma} \geq 15 {\rm ~GeV}$, 
while the ordinary QCD PS was applied to the other processes.

The generated events were passed to PYTHIA 6.425~\cite{Sjostrand:2006za} 
in order to simulate further small-$Q^{2}$ phenomena down to the hadron level.
The PYTHIA simulation was carried out with its default setting, 
except for {\tt parp(67) = 1.0} and {\tt parp(71) = 1.0}, 
as usual~\cite{Odaka:2011hc}.
The kinematical cuts in Eq.~(\ref{eq:selection}) were applied to 
the hadron-level events.
Although the PYTHIA simulation produces some soft photons, 
there is no ambiguity in the selection of the two candidate photons 
since the GR@PPA simulation produces only two photons.
The event generation conditions were set very loose 
in order not to affect the final results.
Note that the $gg \rightarrow \gamma\gamma$ and two-fragmentation 
processes are not included in the current simulation.

\begin{figure}[tp]
\centerline{\psfig{file=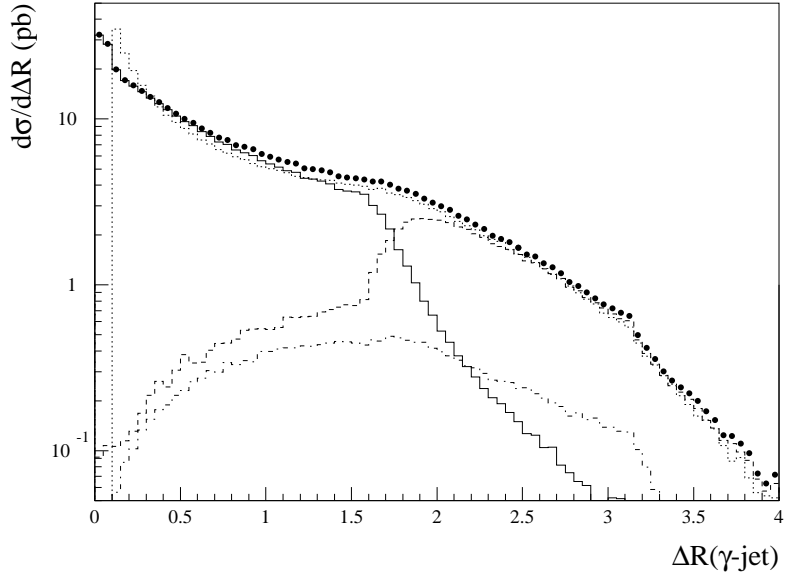,width=120mm}}
\caption{$\Delta R(\gamma\text{-jet})$ distribution of the fully simulated 
radiative processes.
The fragmentation process (solid histogram) and the LLL-subtracted 
$qg \rightarrow \gamma\gamma + q$ process (dashed histogram) are combined to 
a smooth distribution (fulled circles), 
which should be compared with the naive $qg \rightarrow \gamma\gamma + q$ 
ME-prediction (dotted histogram).
The dot-dashed histogram shows the distribution of 
$q\bar{q} \rightarrow \gamma\gamma + g$ events.
\protect\label{fig:draj_full}}
\end{figure}

The resultant $\Delta R(\gamma\text{-jet})$ distributions from the radiative 
processes are shown in Fig.~\ref{fig:draj_full}.
We obtain a smooth distribution plotted with filled circles by combining 
the distributions for the fragmentation process (solid histogram) and 
LLL-subtracted $qg \rightarrow \gamma\gamma + q$ process (dashed histogram).
In the fragmentation process, $\Delta R(\gamma\text{-jet})$ has been derived 
from the parton-level $qg \rightarrow \gamma\gamma + q$ events 
reconstructed as described in a previous section, 
while it has been derived directly from the generated parton-level event 
in the LLL-subtracted $qg \rightarrow \gamma\gamma + q$ process.
The combined result should be compared with the naive 
$qg \rightarrow \gamma\gamma + q$ ME-prediction shown with a dotted histogram, 
as in Fig.~\ref{fig:drgamq}.
We can see that they are in reasonable agreement in most of the regions, 
and that the divergence at $\Delta R(\gamma\text{-jet}) = 0$ 
in the naive ME-prediction is naturally regularized in the combined result.
We can also see a substantial enhancement of the combined result 
with respect to the naive ME-prediction in a wide range. 
This enhancement can be attributed to an effect of multiple QCD radiations 
in the final state. 
Although the distribution is finite, the events are strongly concentrated 
in a small $\Delta R(\gamma\text{-jet})$ region.
The fragmentation photon is likely to be surrounded by hadrons 
when $\Delta R(\gamma\text{-jet})$ is small.
Isolation requirements will suppress these events. 
On the other hand, isolation requirements will not be effective for 
other processes.
The contribution from the $q\bar{q} \rightarrow \gamma\gamma + g$ process is 
illustrated with a dot-dashed histogram in Fig.~\ref{fig:draj_full}.
This distribution does not have any concentration at small 
$\Delta R(\gamma\text{-jet})$.

\begin{figure}[tp]
\centerline{\psfig{file=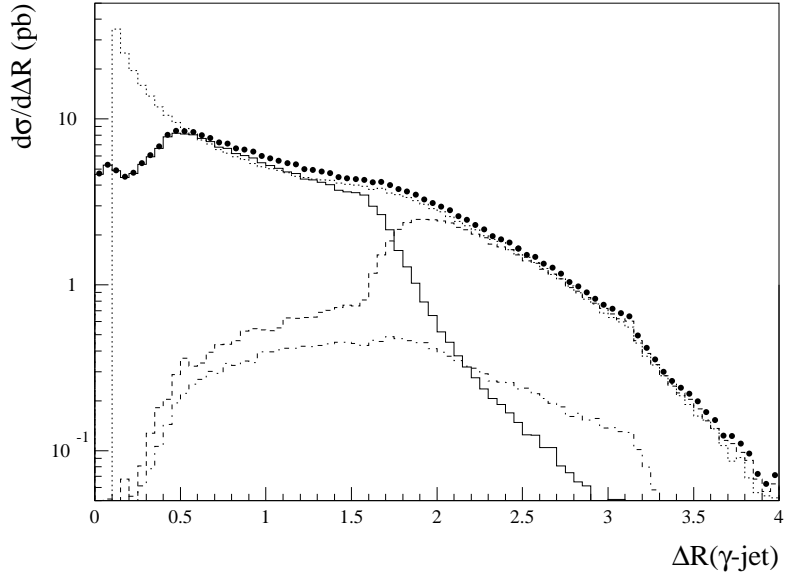,width=120mm}}
\caption{$\Delta R(\gamma\text{-jet})$ distribution of the radiative processes 
after the isolation cut is applied.
The notation of the histograms and plot is the same as that in 
Fig.~\ref{fig:draj_full}.
\protect\label{fig:draj_iso}}
\end{figure}

We impose an isolation condition by using a cone $E_{T}$ 
which is defined at the hadron level as
\begin{equation}\label{eq:etiso}
  E_{T,\text{iso}} = \sum_{\Delta R < R_{\text{iso}} }E_{T},
\end{equation}
where $E_{T}$ is the transverse component of the particle energy 
with respect to the beam direction. 
The sum is taken over all particles inside a given $\Delta R$ cone 
around the photon, excluding the photon itself and neutrinos.
We take as $R_{\text{iso}} = 0.4$ and require that 
$E_{T,\text{iso}} <$ 15 GeV.
The $\Delta R(\gamma\text{-jet})$ distribution after applying the cut 
is shown in Fig.~\ref{fig:draj_iso}.
We can see that the isolation cut strongly suppresses the events 
in $\Delta R(\gamma\text{-jet}) \lesssim 0.4$, as expected.
The events in this region are dominantly produced by the fragmentation 
process.
About 40\% of the events from the fragmentation process were rejected 
by the isolation cut.
On the other hand, the cut has a small impact on the LLL-subtracted 
radiative processes.
Although, about 5\% of the events were rejected, 
including the contribution from $q\bar{q} \rightarrow \gamma \gamma$,
the reduction in the cross section is only 1\% 
because the nearly equal number of positive- and negative-weight events 
exhibited at small $\Delta R(\gamma\text{-jet})$ 
in the LLL-subtracted $qg \rightarrow \gamma \gamma + q$, 
as shown in Fig.~\ref{fig:drgamq}.

\begin{figure}[tp]
\centerline{\psfig{file=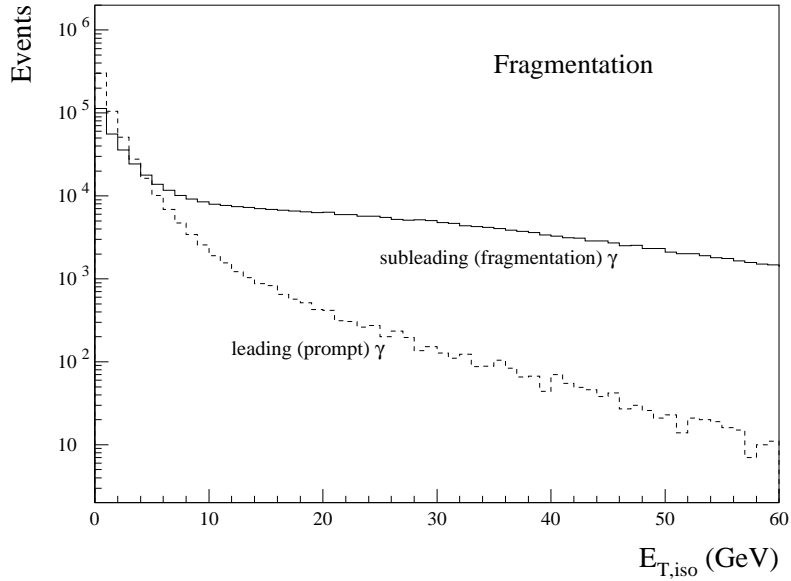,width=120mm}}
\caption{$E_{T{\rm ,iso}}$ distributions for the selected two photons 
in the fragmentation events.
The distribution for the leading (highest-$E_{T}$) photon is shown with 
a dashed histogram and that for the subleading photon with a solid histogram.
\protect\label{fig:etiso}}
\end{figure}

Figure~\ref{fig:etiso} shows the $E_{T{\rm ,iso}}$ distributions for the 
two photons in the fragmentation events.
The distribution for the leading (highest-$E_{T}$) photon is shown with a 
dashed histogram, and that for the subleading photon with a solid histogram.
The leading photon predominantly corresponds to the prompt photon produced 
by the $qg \rightarrow \gamma q$ interaction, 
and the subleading photons are mostly produced in the fragmentation 
of the quark.
Parton showers produce low-energy partons in a wide area 
and the underlying-event simulation is activated in the PYTHIA simulation. 
These hadronic activities produce non-zero $E_{T{\rm ,iso}}$ even for 
the prompt photon by an accidental overlap.
This contribution steeply decreases as $E_{T{\rm ,iso}}$ increases, 
as we can see in the dashed histogram in Fig.~\ref{fig:etiso}.
In contrast, 
the distribution for the fragmentation photons shows a long tail extending 
to large $E_{T{\rm ,iso}}$.
This tail corresponds to the hadronic activity produced in association with 
the photon, {\it i.e.}, the remnant jet.

\begin{figure}[tp]
\centerline{\psfig{file=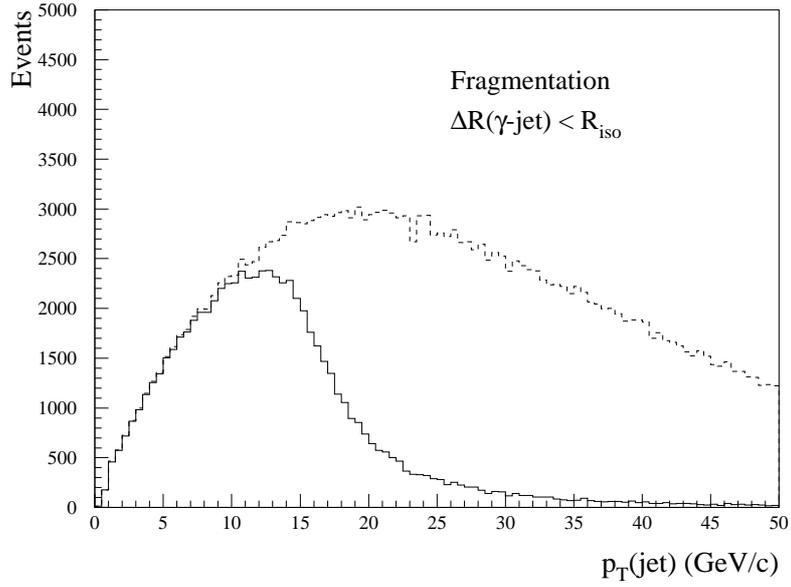,width=120mm}}
\caption{$p_{T}$ distribution of the quark in the parton-level 
$qg \rightarrow \gamma \gamma + q$ events reconstructed 
for the fragmentation events in $\Delta R(\gamma\text{-jet}) < 0.4$.
The dashed histogram shows the distribution before the isolation cut 
and the solid histogram shows that after the cut, 
$E_{T,\text{iso}} <$ 15 GeV.
\protect\label{fig:ptjet}}
\end{figure}

Although small-$\Delta R(\gamma\text{-jet})$ events are strongly suppressed, 
the isolation cut does not produce a sharp edge 
in the $\Delta R(\gamma\text{-jet})$ distribution 
owing to a spread of the remnant jets.
Figure~\ref{fig:ptjet} shows the $p_{T}$ distribution of the remnant quark 
in the parton-level $qg \rightarrow \gamma \gamma + q$ events reconstructed 
for the fragmentation events.
The distribution for those events inside the parton-level isolation cone, 
$\Delta R(\gamma\text{-jet}) < 0.4$, are plotted.
The solid and dashed histograms show the distribution after and before 
the isolation cut, respectively. 
The events with $p_{T}({\rm jet}) > 15 {\rm ~GeV}/c$ are likely to be 
rejected by the cut.
However, the cut does not produce a sharp edge.
On the other hand, a sharp edge should appear in this distribution and 
the distribution in Fig.~\ref{fig:draj_iso} if we apply a similar cut 
at the parton level.
This kind of smearing effect due to the final-state PS and hadronization 
cannot be evaluated without using simulations at the hadron level.

\section{Combined results}

\begin{figure}[tp]
\centerline{\psfig{file=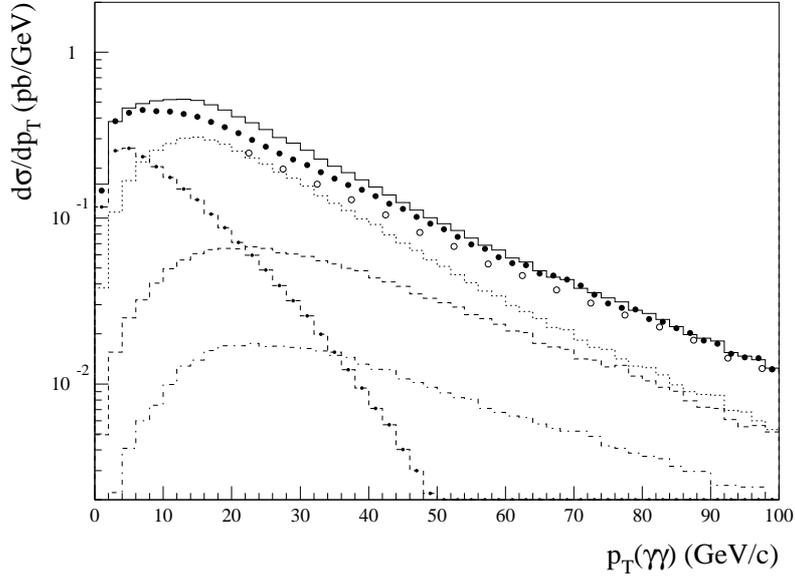,width=120mm}}
\caption{$p_{T}$ distribution of the diphoton ($\gamma\gamma$) system 
in the simulated events satisfying the kinematical condition in 
Eq.~(\ref{eq:selection}) and the isolation condition.
The sum (solid histogram) is presented together with the distributions of the 
subprocesses: $q\bar{q} \rightarrow \gamma \gamma$ (dashed histogram with dots), 
the LLL subtracted $qg \rightarrow \gamma \gamma + q$ (dashed histogram) 
and $q\bar{q} \rightarrow \gamma \gamma + q$ (dot-dashed histogram), 
and the fragmentation process (dotted histogram).
The result is compared with the predictions from RESBOS (filled circles) 
and DIPHOX (open circles).
The DIPHOX prediction is presented at $p_{T} > 20 {\rm ~GeV}/c$ only.
\protect\label{fig:ptgamgam}}
\end{figure}

We can obtain a consistent simulation result by combining the two 
simulations described in the previous section.
Though we have not mentioned to the result from the 
$q\bar{q} \rightarrow \gamma \gamma$ process previously, 
this process is simultaneously simulated with the LLL-subtracted 
radiative processes. 
The initial-state PS applied to the $q\bar{q} \rightarrow \gamma \gamma$ 
events regularizes the initial-state LL components subtracted 
from the radiative processes.

Figure~\ref{fig:ptgamgam} shows the $p_{T}$ distribution of the diphoton 
($\gamma\gamma$) system from the combined simulation. 
The isolation cut described in the previous section has been applied 
to the simulated events selected with the kinematical condition 
in Eq.~(\ref{eq:selection}).
In addition to the combined result (solid histogram), 
the distributions of subprocesses are separately shown. 
We can see that the $p_{T}$ dependence is significantly different between 
the subprocesses.
The integrated contribution from each subprocess is: 
4.2 pb (23\%) from $q\bar{q} \rightarrow \gamma \gamma$, 
9.7 pb (54\%) from the fragmentation, 
and 3.3 pb (18\%) and 0.9 pb (5\%) from the LLL-subtracted 
$qg \rightarrow \gamma \gamma + q$ 
and $q\bar{q} \rightarrow \gamma \gamma + g$ processes, respectively.
It is remarkable that the contribution from the lowest-order process, 
$q\bar{q} \rightarrow \gamma \gamma$, is smaller than 1/4 of the sum.
Furthermore, the $p_{T}(\gamma\gamma)$ spectrum of 
$q\bar{q} \rightarrow \gamma \gamma$ is apparently different 
from the summed spectrum.
In the weak-boson production processes which we have previously studied, 
the production is totally dominated by the lowest-order process 
at small $p_{T}$~\cite{Odaka:2009qf}.
On the other hand, in the diphoton production, 
the lowest-order process is not dominant even at very small $p_{T}$ 
($\lesssim 10 {\rm ~GeV}/c$) as we can see in Fig.~\ref{fig:ptgamgam}.
Studies with the lowest-order process alone are not sufficient 
even if large-$p_{T}$ contributions are intensively excluded.
We can also see that simulations are not sufficient 
even if we include the fragmentation process in addition to 
the lowest-order process.
The contribution from the LLL-subtracted $qg \rightarrow \gamma \gamma + q$ 
process, which supplements the fragmentation process in the hard-radiation 
region, is nearly comparable to the lowest-order process, 
and has a $p_{T}(\gamma \gamma)$ spectrum remarkably different from 
the lowest-order and fragmentation processes.

It must be noted that the composition described in the above 
is not physically meaningful.
The above result is obtained when we separate the soft and hard radiations 
at $p_{T}$ of the $q\bar{q} \rightarrow \gamma \gamma$ or 
$qg \rightarrow \gamma q$ interaction.
The composition changes if we adopt other definitions.
Incidentally, it must also be noted that the fragmentation process is 
a part of the $qg \rightarrow \gamma \gamma + q$ process.
The fragmentation process has been introduced to regularize the final-state 
divergence in $qg \rightarrow \gamma \gamma + q$.
The sum of the fragmentation and LLL-subtracted 
$qg \rightarrow \gamma \gamma + q$ processes amounts to more than 70\% 
of the total yield in our simulation.
This fraction would not change dramatically even if we moved the separation 
criteria in a reasonable range.
The domination of the $qg \rightarrow \gamma \gamma + q$ process is 
a characteristic feature of the diphoton production in hadron collisions, 
at least, at the LHC with a typical Higgs-search condition.

The simulated $p_{T}(\gamma\gamma)$ spectrum is compared with the prediction 
from RESBOS~\cite{Balazs:2007hr} in Fig.~\ref{fig:ptgamgam}, 
for which we have used CTEQ6M~\cite{Pumplin:2002vw} for PDF 
and required the same kinematical and isolation conditions.
RESBOS provides an NLO prediction 
in which initial-state soft QCD radiations are resummed. 
It is considered to be most reliable at present as far as the 
$p_{T}(\gamma\gamma)$ distribution is concerned.
Unfortunately, since the resummation can only predict inclusive properties, 
RESBOS cannot provide the hadron-level information.
Thus, the selections are all applied at the parton level.
The contribution from the $gg \rightarrow \gamma\gamma$ process 
is not included in the compared RESBOS prediction 
because this process is yet to be included in our simulation.
The RESBOS prediction (filled circles) shown in Fig.~\ref{fig:ptgamgam} 
is smaller than the result presented in the original paper~\cite{Balazs:2007hr} 
because the $m_{\gamma\gamma}$ constraint in Eq.~(\ref{eq:selection}) 
is additionally required.
We have confirmed that we can obtain a prediction close to the original result 
if the $m_{\gamma\gamma}$ constraint is removed.
As we can see in Fig.~\ref{fig:ptgamgam}, 
our simulation gives a substantially larger cross section than 
the RESBOS prediction although the overall tendency is consistent.
The total cross section is 15.6 pb from RESBOS and 18.1 pb from our simulation.

The open circles in Fig.~\ref{fig:ptgamgam} show the prediction from 
DIPHOX~\cite{Binoth:1999qq} with the CTEQ6M PDF. 
DIPHOX is a full NLO calculation for diphoton production. 
The predictions are considered to be comparable with those from RESBOS.
However, because soft QCD effects are not resummed, 
it cannot provide a continuous spectrum at small $p_{T}(\gamma \gamma)$.
In Fig.~\ref{fig:ptgamgam}, we have plotted the DIPHOX prediction 
in a relatively large $p_{T}(\gamma \gamma)$ region ($> 20 {\rm ~GeV}/c$) 
where we can obtain a continuous $p_{T}(\gamma \gamma)$ prediction.
We have required the kinematical condition in Eq.~(\ref{eq:selection})  
and the same isolation condition. 
The plotted prediction is the sum of the results for the "direct" process 
and the one-fragmentation process separately evaluated.
We have used the LO prediction for the one-fragmentation process, 
and subtracted the $gg \rightarrow \gamma \gamma$ contribution 
from the direct process.
The obtained DIPHOX prediction is continuously smaller than our simulation 
and the RESBOS prediction in the displayed $p_{T}(\gamma \gamma)$ range, 
although the difference becomes smaller as $p_{T}(\gamma \gamma)$ 
becomes larger. 
This $p_{T}(\gamma \gamma)$ behavior is reasonable because DIPHOX does 
not include initial-state radiation effects.
The total cross section from this DIPHOX prediction is 13.7 pb.

\begin{figure}[tp]
\centerline{\psfig{file=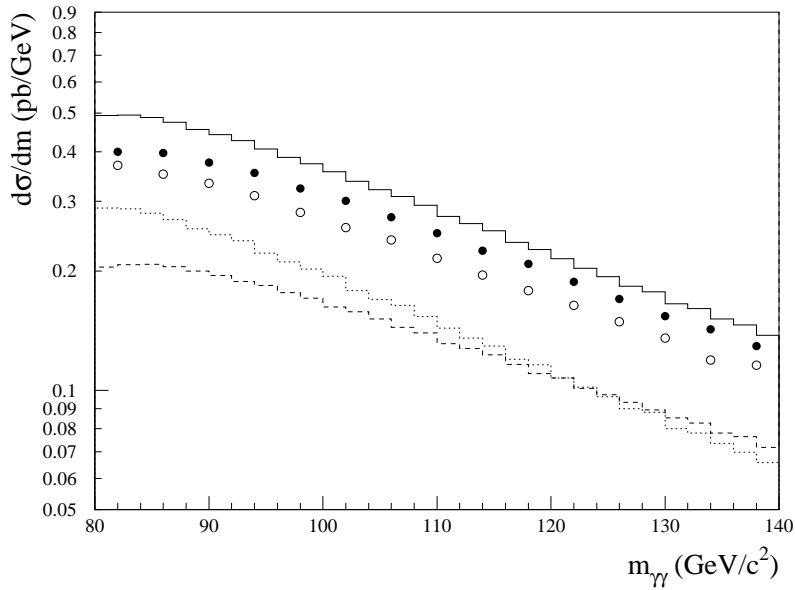,width=120mm}}
\caption{Invariant-mass ($m_{\gamma\gamma}$) distribution 
of the diphoton system.
Together with the combined result (solid histogram), 
the contribution from the fragmentation process (dotted histogram) and 
that from the other processes (dashed histogram) are separately presented.
The result is compared with the predictions from RESBOS (filled circles) 
and DIPHOX (open circles).
\protect\label{fig:mgamgam}}
\end{figure}

Figure~\ref{fig:mgamgam} compares the diphoton invariant-mass 
($m_{\gamma\gamma}$) distribution obtained from our simulation (solid histogram) 
with those from RESBOS (filled circles) and DIPHOX (open circles).
The results are almost comparable to each other, 
although there are substantial overall shifts between them and 
small differences in the $m_{\gamma\gamma}$ dependence. 
If we ignore the overall shifts, 
the DIPHOX prediction is closer to our simulation than RESBOS.
The small-$m_{\gamma\gamma}$ fraction tends to be larger in our simulation 
and DIPHOX.
Here, we have to remind that the isolation requirement in RESBOS is
markedly different from that in DIPHOX and our simulation.
Our simulation sums up the $E_{T}$ of particles inside the isolation cone, 
$\Delta R < 0.4$,
and rejects events if the summed $E_{T}$, $E_{T,{\rm iso}}$, exceeds 
the threshold of 15 GeV.
In principle, DIPHOX does the same rejection at the parton level.
On the other hand, although the selection condition is literally identical, 
RESBOS rejects the whole LL contribution 
when $\Delta R(\gamma\text{-jet}) < 0.4$, 
in order to regularize the final-state divergence.
Events with $E_{T,{\rm iso}} < 15 {\rm ~GeV}$ are allowed only for 
non-LL components.
As we can see in Fig.~\ref{fig:draj_full},
the fragmentation process that simulates the LL component in our simulation 
dominate the cross section inside the isolation cone.
The contribution from the fragmentation process and that from the other 
processes are separately shown with dotted and dashed histograms, 
respectively, in Fig.~\ref{fig:mgamgam}.
We can see that the fragmentation process shows a substantial enhancement 
at smaller $m_{\gamma\gamma}$ with respect to the others, 
suggesting that the small-$m_{\gamma\gamma}$ enhancement in DIPHOX 
and our simulation may be explained by the difference 
in the isolation requirement.
The difference between our simulation and RESBOS, 
which we can see in Fig.~\ref{fig:ptgamgam}, 
seems to support this argument; the difference is 
large in the region where the fragmentation contribution is large.

\begin{figure}[tp]
\centerline{\psfig{file=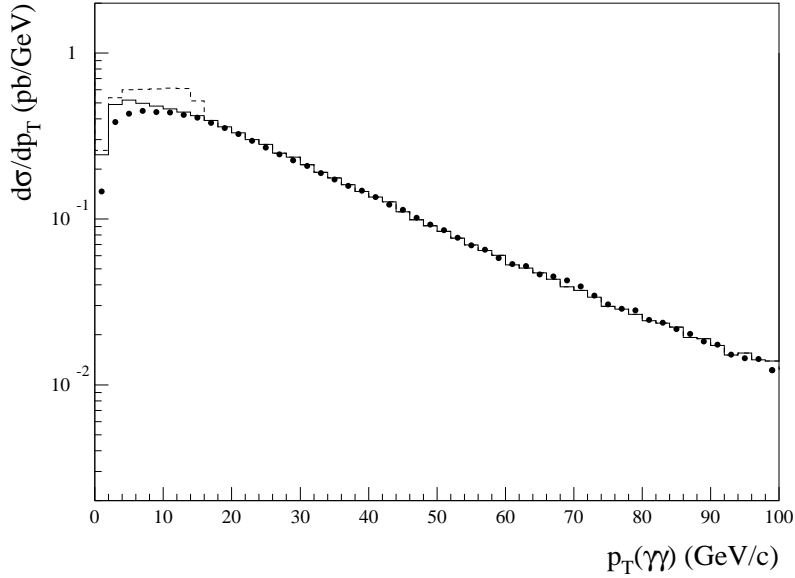,width=120mm}}
\caption{$p_{T}$ distribution of the diphoton ($\gamma\gamma$) system 
obtained from the hybrid simulation. 
The solid histogram shows the result from a RESBOS-style simulation 
in which all fragmentation events with $\Delta R(\gamma\text{-jet}) < 0.4$ 
are rejected.
The dashed histogram has been obtained from a simulation 
in which the standard isolation cut, $E_{T,{\rm iso}} < 15 {\rm ~GeV}$, 
is applied at the parton level in all radiative processes.
The results are compared with the RESBOS prediction shown with filled circles.
\protect\label{fig:ptgamgam2}}
\end{figure}

As a test, we have carried out a hybrid simulation in order to clarify 
the reason for the observed difference. 
In this simulation, we applied the kinematical cut and the isolation 
cut directly to the generated parton-level events 
in the LLL-subtracted radiative processes, 
without applying PS and the PYTHIA simulation.
In the fragmentation process, only the final-state PS was applied to 
the generated events in order to reconstruct the parton-level 
$qg \rightarrow \gamma \gamma + q$ events.
The kinematical cuts were applied to these parton-level events, 
and all events having $\Delta R(\gamma\text{-jet}) < 0.4$ were rejected 
in order to simulate the calculation in RESBOS.
These are extensions of the simulations in the matching test.
On the other hand, the full simulation down to the hadron level was 
applied to the lowest-order $q\bar{q} \rightarrow \gamma \gamma$ process 
in order to make the combined result comparable with the resummed calculation 
in RESBOS.

The combined $p_{T}(\gamma \gamma)$ distribution from the hybrid simulation 
is presented with a solid histogram in Fig.~\ref{fig:ptgamgam2}, 
to be compared with the RESBOS prediction plotted with filled circles.
We can see a complete agreement between them over almost the entire 
$p_{T}(\gamma \gamma)$ region, 
except for a narrow region at small $p_{T}(\gamma \gamma)$.
The total cross section from the hybrid simulation is 16.5 pb, 
which is only 6\% larger than the RESBOS prediction.
The agreement is remarkable but not surprising 
because the radiative processes which dominate the total cross section 
are evaluated at the tree level even in NLO calculations, 
such as RESBOS and DIPHOX.
The contribution of non-divergent soft/virtual corrections which are 
missing in our simulation must be small, since the contribution of 
the lowest-order process to which the corrections are to be applied 
is small.
The difference at small $p_{T}(\gamma \gamma)$ may have been caused 
by the difference between the PS and the resummation.
In any case, the agreement implies that the calculations are reasonably 
carried out both in RESBOS and our simulation.

The dashed histogram in Fig.~\ref{fig:ptgamgam2} shows the result 
from another hybrid simulation, 
in which we allowed those events with $E_{T,{\rm iso}} < 15 {\rm ~GeV}$ 
in the fragmentation process.
This change resulted in a step-like structure in the $p_{T}(\gamma \gamma)$ 
spectrum.
The reason for this is quite simple.
The $p_{T}$ of the diphoton system is equal to the $p_{T}$ of the remnant jet 
in the fragmentation process, 
and $E_{T,{\rm iso}}$ is equal to this $p_{T}$ if the remnant jet is inside 
the isolation cone.
Therefore, the allowance of non-zero $E_{T,{\rm iso}}$ 
produces an enhancement only at $p_{T}(\gamma \gamma) < 15 {\rm ~GeV}$.
This change increased the total cross section to 18.1 pb, 
which is in good agreement with the full simulation result.
This step-like enhancement can be considered as the source of the difference 
between our simulation and RESBOS in Fig.~\ref{fig:ptgamgam}.
The enhancement must have been smeared by multiple QCD-radiation in PS and 
hadronization.
Hence, the difference in Fig.~\ref{fig:ptgamgam} must be reasonable and 
shows the effects that are not supported in RESBOS.

\begin{figure}[tp]
\centerline{\psfig{file=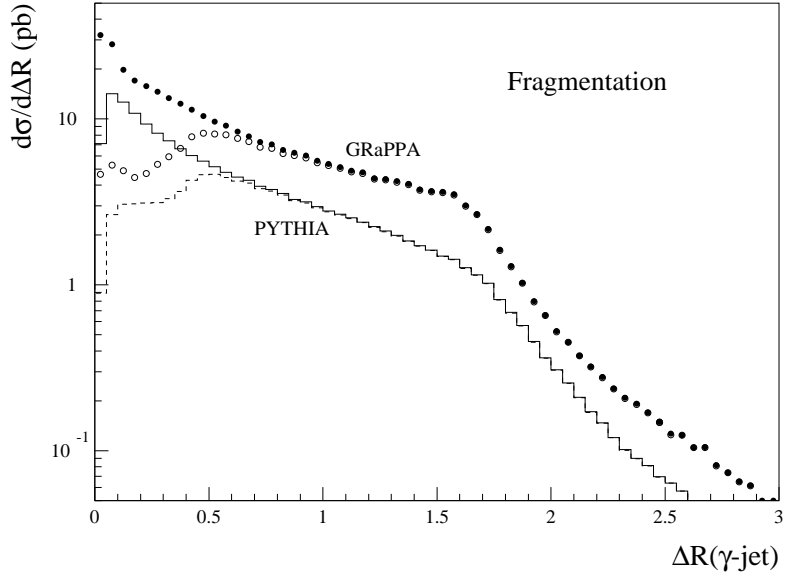,width=120mm}}
\caption{$\Delta R(\gamma\text{-jet})$ distribution of the fragmentation 
process simulated with the PYTHIA PS and our PS in the standard setting.
The distribution before and after the isolation cut is shown with a 
solid and dashed histograms, respectively, for the result with the PYTHIA PS.
The corresponding distributions with our PS are plotted with filled 
circles and open circles, respectively.
\protect\label{fig:pythia}}
\end{figure}

We have carried out a study by using our custom-made PS in the above.
We performed a similar study previously~\cite{Odaka:2011qg} 
by using the PYTHIA PS~\cite{Sjostrand:2006za} 
for generating the fragmentation events.
The matching properties were not fully satisfactory and the resultant 
diphoton-production cross section was smaller than 
the RESBOS prediction in that study.
These were mainly caused by the fact that the photon yield from the PYTHIA PS 
was significantly smaller than that from our PS.
The $\Delta R(\gamma\text{-jet})$ distributions obtained with 
the PYTHIA PS and our PS are compared in Fig.~\ref{fig:pythia}.
We can see that the PYTHIA-PS predictions are almost always 
smaller than ours by about a factor of two.
We do not understand the reason for this difference.
In addition, though it may be too much detail, 
the drop in the first bin that we can see in the PYTHIA results is caused 
by the absence of small-$Q^{2}$ photon radiations ($Q < 1 {\rm ~GeV}$). 
Such small-$Q^{2}$ radiations are simulated with the FF radiation in our PS.

\section{Conclusion}

We have developed an exclusive event generator for non-resonant 
QED diphoton ($\gamma\gamma$) production at hadron collisions, 
consistently including additional one-jet production. 
The $qg \rightarrow \gamma \gamma + q$ process to be included as a radiative 
process has a final-state QED divergence 
together with an initial-state QCD divergence.
We have developed a final-state subtraction method 
by extending the method developed for initial-state QCD divergences 
(the LLL subtraction). 
The subtraction works as well as we expected.
The differential cross section becomes finite after the subtraction 
and converges to zero at the limit where the original cross section diverges.

We have also developed a final-state parton shower (PS) for generating 
fragmentation events.
The PS implements QED photon radiations as well as ordinary QCD radiations,
and can enforce hard-photon radiation 
in order to improve the generation efficiency.
Small-$Q^{2}$ photon radiations which are not covered by the PS are simulated 
by using a fragmentation function (FF).
The divergent components in the final state which are subtracted 
from the radiative processes are restored with an appropriate regularization 
by applying this PS to $qg \rightarrow \gamma q$ events.
The subtraction and regularization of initial-state QCD divergences are 
carried out in the same way as applied to weak-boson productions.
The generated fragmentation events show a complete matching with the 
LLL-subtracted $qg \rightarrow \gamma \gamma + q$ events.

A consistent simulation sample has been obtained 
by combining the fragmentation events 
with the $q\bar{q} \rightarrow \gamma \gamma$ and LLL-subtracted
$q\bar{q} \rightarrow \gamma \gamma + g$ and 
$qg \rightarrow \gamma \gamma + q$ events 
simultaneously generated in GR@PPA.
The event generation was tested for the LHC design condition, 
$pp$ collisions at 14 TeV. 
A typical Higgs-boson search condition was required to the events simulated 
down to the hadron level.
A typical isolation cut was also applied to the photons.
We observed a reasonable suppression of the events in the collinear region, 
together with a visible smearing in physical distributions 
due to the application of PS and hadronization.

In our simulation, 
for which we have chosen all the energy scales to be equal to $p_{T}$ of 
the non-radiative $q\bar{q} \rightarrow \gamma \gamma$ or 
$qg \rightarrow \gamma q$ events, 
the contribution from the lowest-order $q\bar{q} \rightarrow \gamma \gamma$ 
process is smaller than 1/4 of the total yield.
The contribution from the LLL-subtracted $qg \rightarrow \gamma \gamma + q$ 
process, which supplements the fragmentation process in hard-radiation regions, 
is comparable to it.
A consistent inclusion of this process is necessary for reliable studies. 
If we combine the contributions from the fragmentation and 
LLL-subtracted $qg \rightarrow \gamma \gamma + q$ processes, 
the total $qg \rightarrow \gamma \gamma + q$ contribution amounts to 
more than 70\%.

The combined simulation sample shows a behavior which reasonably agrees 
with the predictions from next-to-leading order (NLO) calculations 
by RESBOS and DIPHOX.
Observed differences can be attributed to the different implementation of the 
isolation requirement, and higher-order QCD and hadronization effects 
which are not supported in NLO calculations. 
It must be noted that the two-fragmentation and $gg \rightarrow \gamma \gamma$ 
and their higher orders are yet to be included in our simulation.

\section*{Acknowledgments}

This work has been carried out as an activity of the NLO Working Group, 
a collaboration between the Japanese ATLAS group and the numerical analysis 
group (Minami-Tateya group) at KEK.
The authors wish to acknowledge useful discussions with the members.


\begin{thebibliography}{99}

\bibitem{ATLAS:2012sk}
ATLAS Collaboration, {\it {Search for the Standard Model Higgs boson in
  the diphoton decay channel with 4.9 $fb^{-1}$ of pp collisions at $\sqrt{s}$
  = 7 TeV with ATLAS}},  arXiv:1202.1414 [hep-ex].

\bibitem{CMS:2012tw}
CMS Collaboration, {\it {Search for the standard model Higgs boson
  decaying into two photons in pp collisions at $\sqrt{s}$ = 7 TeV}},
  arXiv:1202.1487 [hep-ex].

\bibitem{Binoth:1999qq}
T.~Binoth, J.~P. Guillet, E.~Pilon, and M.~Werlen, {\it {A full next-to-leading
  order study of direct photon pair production in hadronic collisions}}, 
  Eur. Phys. J. C {\bf 16} (2000) 311;
  arXiv:hep-ph/9911340.

\bibitem{Balazs:2007hr}
C.~Balazs, E.~L. Berger, P.~M. Nadolsky, and C.~P. Yuan, {\it {Calculation of
  prompt diphoton production cross sections at Tevatron and LHC energies}},
  Phys. Rev. D {\bf 76} (2007) 013009;
  arXiv:0704.0001 [hep-ph].

\bibitem{Hoeche:2009xc}
S.~Hoeche, S.~Schumann, and F.~Siegert, {\it {Hard photon production and
  matrix-element parton-shower merging}},  Phys. Rev. D {\bf 81} (2010)
  034026; arXiv:0912.3501 [hep-ph].

\bibitem{Bahr:2008pv}
M.~Bahr {\em et~al.}, {\it {Herwig++ Physics and Manual}},  
  Eur. Phys. J. C {\bf 58} (2008) 639; arXiv:0803.0883 [hep-ph].

\bibitem{D'Errico:2011sd}
L.~D'Errico and P.~Richardson, {\it {Next-to-Leading-Order Monte Carlo
  Simulation of Diphoton Production in Hadronic Collisions}},
  arXiv:1106.3939 [hep-ph].

\bibitem{Odaka:2011hc}
S.~Odaka and Y.~Kurihara, {\it {GR@PPA 2.8: initial-state jet matching for
  weak-boson production processes at hadron collisions}},  Comput. Phys.
  Commun. {\bf 183} (2012) 1014;
  arXiv:1107.4467 [hep-ph].

\bibitem{Odaka:2012da}
S.~Odaka, {\it {GR@PPA 2.8.3 update}}, arXiv:1201.5702 [hep-ph].

\bibitem{Ishikawa:1993qr}
Minami-Tateya Group, T.~Ishikawa {\em et~al.}, {\it {GRACE
  manual: Automatic generation of tree amplitudes in Standard Models: Version
  1.0}}, KEK Report KEK-92-19 (1993).

\bibitem{Kurihara:2002ne}
Y.~Kurihara {\em et~al.}, {\it {QCD event generators with next-to-leading order
  matrix- elements and parton showers}},  Nucl. Phys. {\bf B654} (2003) 301; 
  arXiv:hep-ph/0212216.

\bibitem{Odaka:2007gu}
S.~Odaka and Y.~Kurihara, {\it {Initial-state parton shower kinematics for NLO
  event generators}},  Eur. Phys. J. C {\bf 51} (2007) 867;
  arXiv:hep-ph/0702138.

\bibitem{Odaka:2009qf}
S.~Odaka, {\it {Simulation of $Z$ boson $p_{T}$ spectrum at Tevatron by
  leading-order event generators}},  Mod. Phys. Lett. A {\bf 25} (2010) 3047; 
  arXiv:0907.5056 [hep-ph].

\bibitem{Sjostrand:2006za}
T.~Sjostrand, S.~Mrenna, and P.~Z. Skands, {\it {PYTHIA 6.4 Physics and
  Manual}},  JHEP {\bf 05} (2006) 026; arXiv:hep-ph/0603175.

\bibitem{Pumplin:2002vw}
J.~Pumplin {\em et~al.}, {\it {New generation of parton distributions with
  uncertainties from global QCD analysis}},  JHEP {\bf 07} (2002) 012;
  arXiv:hep-ph/0201195.

\bibitem{Tsuno:2006cu}
S.~Tsuno, T.~Kaneko, Y.~Kurihara, S.~Odaka, and K.~Kato, {\it {GR@PPA 2.7 event
  generator for $pp/p\bar{p}$ collisions}},  Comput. Phys. Commun. {\bf 175} 
  (2006) 665; arXiv:hep-ph/0602213.

\bibitem{Kawabata:1985yt}
S.~Kawabata, {\it {A New Monte Carlo Event Generator for High-Energy Physics}},
   Comput. Phys. Commun. {\bf 41} (1986) 127.

\bibitem{Kawabata:1995th}
S.~Kawabata, {\it {A new version of the multidimensional integration and event
  generation package BASES/SPRING}},  Comp. Phys. Commun. {\bf 88} (1995)
  309.

\bibitem{Bourhis:1997yu}
L.~Bourhis, M.~Fontannaz, and J.~Guillet, {\it {Quarks and gluon fragmentation
  functions into photons}},  Eur. Phys. J. C {\bf 2} (1998) 529;
  arXiv:hep-ph/9704447.

\bibitem{Odaka:2011qg}
S.~Odaka, {\it {A consistent event generation for non-resonant diphoton
  production at hadron collisions}},
  arXiv:1105.5847 [hep-ph].

\end{thebibliography}

\end{document}